\documentclass{elsart}

\usepackage{amssymb}
\usepackage{graphicx}
\usepackage{subfigure}
\usepackage{amsmath}
\usepackage{wasysym}
\usepackage{lineno}

\linenumbers

\begin{document}

\begin{frontmatter}


\title{Monte Carlo Studies of Geomagnetic Field Effects on the Imaging Air
  Cherenkov Technique for the MAGIC Telescope Site} 

\author[label1]{S.C. Commichau\corauthref{COR}},
\corauth[COR]{Corresponding author. Tel.: +41-44-63 32185; fax.: +41-44-63 31104}
\ead{sebastian.commichau@phys.ethz.ch}
\author[label1]{A. Biland},
\author[label2]{J.L. Contreras},
\author[label2]{R. de los Reyes},
\author[label3]{A. Moralejo},
\author[label4,label5]{J. Sitarek},
\author[label5]{D. Sobczy\'{n}ska}
\author{on behalf of the MAGIC collaboration\thanksref{MAGIC}}
\address[label1]{ETH Zurich, CH-8093 Switzerland}
\address[label2]{Universidad Complutense, E-28040 Madrid, Spain}
\address[label3]{Institut de F\'\i sica d'Altes Energies, Edifici Cn., E-08193 Bellaterra (Barcelona), Spain}
\address[label4]{Max-Planck-Institut f\"ur Physik, D-80805 M\"unchen, Germany}
\address[label5]{University of \L \'od\'z, PL-90236 Lodz, Poland}

\thanks[MAGIC]{\textit{URL:} \texttt{http://wwwmagic.mppmu.mpg.de/collaboration/members/index.html}}

\begin{abstract}
Imaging air Cherenkov telescopes (IACTs) detect the Cherenkov light
from extensive air showers (EAS) initiated by very high energy (VHE)
$\gamma$-rays impinging on the Earth's atmosphere. Due to the overwhelming
background from hadron induced EAS, the discrimination of the rare
$\gamma$-like events is vital.
The influence of the geomagnetic field (GF) on the development
of EAS can further complicate the imaging air Cherenkov technique.
The amount and the angular distribution of Cherenkov light
from EAS can be obtained by means of Monte Carlo (MC) simulations.
Here we present the results from dedicated MC studies of GF effects on
images from $\gamma$-ray initiated EAS for the MAGIC telescope
site, where the GF strength is $\sim 40\,\mu\text{T}$. The results from the
MC studies suggest that GF effects degrade not only measurements of very low energy
$\gamma$-rays below $\sim 100\,\text{GeV}$ but also those at TeV-energies.
\end{abstract}


\end{frontmatter}

\section{Introduction}

Imaging air Cherenkov telescopes (IACTs) aim at the detection of Cherenkov light
from extensive air showers (EAS) initiated by very high energy (VHE)
$\gamma$-rays impinging on the Earth's atmosphere.
IACTs make use of the differences between the angular distributions of
Cherenkov light from $\gamma$-ray and hadron induced EAS to efficiently discriminate the
hadronic background. Due to the overwhelming abundance of
hadrons in the cosmic rays (mostly protons), the discrimination of the rare
$\gamma$-ray events is rather difficult, in particular for energies
below 100\,GeV. Only a small fraction of the recorded data are due to
$\gamma$-ray initiated EAS.
The influence of the geomagnetic field (GF) on the development
of EAS can further complicate the suppression of the hadronic background and
therefore reduces the sensitivity of an instrument.
In addition, GF effects can systematically affect
the energy resolution of an IACT \cite{com0701,wen04,rey08}.\\
The influence of the GF on EAS was already qualitatively discussed in 1953
\cite{coc54}. It was pointed out that the east-west separation of electrons and positrons in EAS
due to the GF can be non-negligible compared to the
displacement from multiple Coulomb scattering.
Furthermore, the Lorentz force systematically deflects the particles into opposite directions whereas the
displacement due to multiple Coulomb scattering is random. It was also argued that
GF effects are relatively less important for hadron induced EAS than
for $\gamma$-ray induced EAS.
The scattering angles occurring in nuclear interactions of hadronic EAS give
rise to a lateral displacement of the shower particles much larger than that due to the influence of
the GF. The comparatively large transverse momenta of the secondary particles result in a large angular spread of the
directions of the electromagnetic sub-cascades generated by pion decay.\\
The influence of the GF on the average lateral spread of atmospheric Cherenkov
radiation was studied by means of computer simulations already more than 30 years ago
\cite{por73}. GF effects on real
$\gamma$-ray and proton initiated EAS were later on studied using a non-imaging Cherenkov
telescope \cite{bow92}. It was reported that the influence of the GF on proton
initiated EAS results in a significant reduction of the count rate.
It was shown elsewhere \cite{lan94} that IACT measurements of TeV $\gamma$-rays from the Crab
nebula were not significantly affected when the component of the GF normal to
the shower axis, i.e. transversal
component of the GF, was below $35\,\mu\text{T}$. The average shape and reconstructed intensity of Cherenkov
images from hadrons was found to be independent of the transversal
component of the GF. However, it was pointed out that the instrument was not
sensitive enough to study GF effects.
More recent measurements of $\gamma$-ray showers carried out with a
transversal component of the GF strength of $|\vec{B}_\perp|> 40\,\mu\text{T}$
revealed GF effects in observational data compatible to those predicted by MC
simulations, both for $\gamma$-ray and hadron showers
\cite{cha9901,cha9902}. The authors suggest that
for EAS developing under unfavourable orientation with
respect to the direction of the GF the corresponding Cherenkov light images in
the camera of an IACT will be rotated. As the information on the orientation of shower images
provides the most powerful discrimination between $\gamma$-ray shower images
from a point-like source
and any unwanted isotropic background (mainly due to hadrons) this results in a degradation of the
sensitivity of an IACT. However, it was demonstrated that a correction for GF effects
in $\gamma$-ray initiated Cherenkov images is possibly resulting in an
increased detection significance and better sensitivity of the IACT
\cite{cha00,aye01}. The correction for GF effects required simulated
$\gamma$-ray showers.\\
IACTs currently in operation offer improved imaging capabilities, i.e. better optical point spread function
(PSF), pixel resolution and timing capabilities of the electronics, and are
therefore more sensitive to GF effects than previous instruments.\\
In this paper we present results from dedicated MC studies of GF effects
on the imaging technique. The studies were carried out for the Major Atmospheric Gamma-ray Imaging
Cherenkov (MAGIC) telescope \cite{lor04,bar98}, which is located on the Canary Island
of La Palma at the Roque de los Muchachos Observatory at 2200\,m altitude ($28.45^\circ$\,N,$17.54^\circ$\,W).

\section{The MAGIC Telescope}

The 17\,m diameter MAGIC telescope
is currently the largest single dish IACT in operation.
The imaging camera in the focal plane of the tessellated parabolic reflector
consists of 577 photomultiplier tubes (PMTs), all of which are arranged in a
hexagonal configuration. The inner part of the
camera is equipped with 397 PMTs of a diameter of $0.1^\circ$ whereas the outer
part is equipped with larger PMTs of a diameter of $0.2^\circ$.
The reflector has a focal length of 17\,m.
The field of view of the camera is $3.5^\circ$ and
the angular resolution for $\gamma$-rays is about $0.1^\circ$, depending on the energy.
The telescope is in continuous operation since summer 2004. It allows
for a detection of a $\gamma$-ray source with an absolute intensity of $\sim
2\,\%$ of the Crab nebula and similar energy spectrum within 50\,hours at energies
$>200$\,GeV on a significance level of 5 standard deviations.\\
MAGIC is currently being upgraded through the addition of a twin telescope to achieve
an improved sensitivity and a lower energy threshold \cite{goe07}. Further
technical details and information on the performance of the instrument can be
found elsewhere \cite{alb0701}.

\clearpage

\section{The Geomagnetic Field at the MAGIC Site}

The component of the GF normal to the shower axis is relevant for the east-west
separation of electrons and positrons during the shower development.
For this study the telescope optical axis has always been set parallel to the
direction of the primary $\gamma$-ray.

\begin{figure}[!h] \centering
  \subfigure[]{\includegraphics[scale=.34]{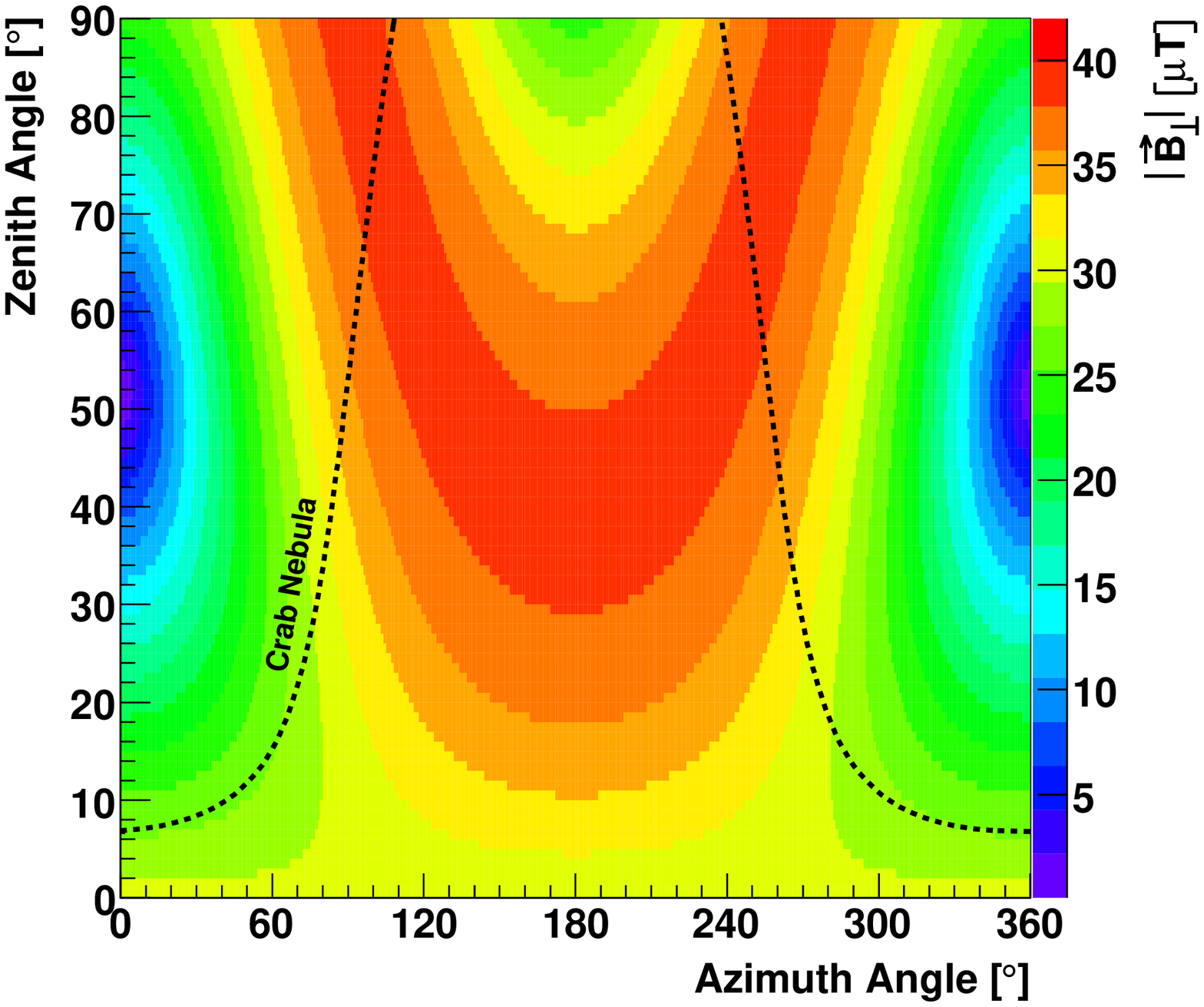}}\qquad
  \subfigure[]{\includegraphics[scale=.45]{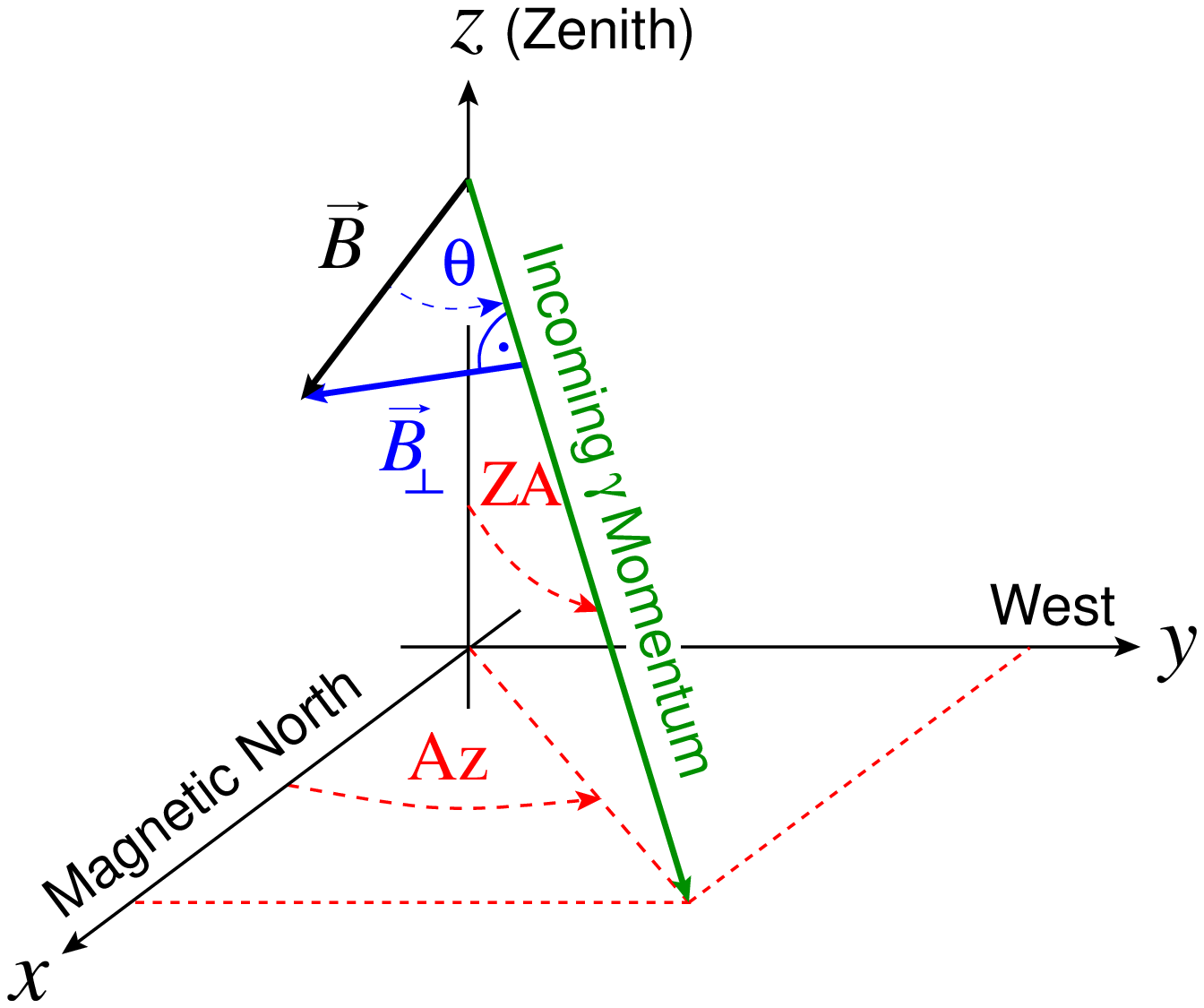}}
  \caption{(a) The absolute value of the component of the GF normal to the direction of
  the EAS versus azimuth (Az) angle and zenith angle (ZA) for the Roque de los
  Muchachos observatory on La Palma.
  (b) The definition of the coordinate
  system used throughout this work. $\theta$ denotes the angle between the direction of the EAS
  and the direction of the GF. The Az angle is defined like in the CORSIKA
  program \cite{hec98}, i.e. it refers to the momentum of the incoming
  $\gamma$-ray and is counted counterclockwise from the positive $x$-axis
  towards west. The telescope optical axis has always been set parallel to the
  direction of the primary $\gamma$-ray.}
  \label{figure1}
\end{figure}

Figure \ref{figure1} (a) shows the absolute value of
the GF component $|\vec{B}_\perp|$ normal to the direction of the EAS
versus azimuth (Az) angle and zenith angle (ZA)
for the MAGIC telescope site. The value was determined for 10\,km
a.s.l. according to the epoch 2005 International Geomagnetic Reference Field
(IGRF) model \cite{ngdc}. The MAGIC telescope is focused to a
distance of 10\,km a.s.l., which is the most likely location of the shower
maximum for 100\,GeV $\gamma$-ray induced EAS at small ZAs.
The trajectory of the strongest source
of steady VHE $\gamma$-ray emission in the Galaxy, the Crab nebula, is
indicated.
As the magnetic field lines at La Palma are tilted by $\sim 7^\circ$ westwards
with respect to the meridian \cite{ngdc} the trajectory is asymmetric with respect to
$180^\circ$ Az angle.\\
For La Palma, the minimum influence of the GF is expected to occur
for EAS developing in direction of the magnetic north at $\text{ZA} =(90^\circ-I)\approx
51^\circ$ and $\text{Az}=0^\circ$, where the angle $\theta$ between the shower axis and the GF becomes
smallest (see figure \ref{figure1} (b)), i.e. for EAS developing along the field lines.
$I$ denotes the angle under which the GF lines dip into the Earth's
surface, which is $\sim 39^\circ$ for La Palma \cite{ngdc}. Hence, the maximum influence
is expected for EAS developing perpendicular to the direction of the GF lines,
i.e. for $\text{ZA} \approx 39^\circ$ and $\text{Az}=180^\circ$.\\
The results from the MC studies on the GF effects presented here are specific to the MAGIC
telescope because of the local GF strength and the relevant telescope parameters
like the reflector area, camera pixelisation and the $\gamma$-ray PSF.

\section{Monte Carlo Simulations}

The production of Monte Carlo (MC) data for MAGIC involves three steps \cite{maj05}:

\begin{itemize}
\item[1.] The CORSIKA (COsmic Ray SImulations for KAscade) MC program (version 6.019)
  \cite{hec98} is used to simulate the development of
  $\gamma$-ray and hadron induced extensive air showers (EAS) and the
  production of Cherenkov light for a given set of input parameters,
  like the primary $\gamma$-ray or hadron energy, height above sea level,
  the magnitude and direction of the GF, and so on.
  The GF components are set to the values for the location of
  the MAGIC telescope (La Palma, $28.8^\circ$\,N, $17.9^\circ$\,W) which are provided
  by the IGRF model \cite{ngdc}. The $x$-axis of the Cartesian CORSIKA reference frame is aligned
  with the magnetic north pole and the $y$-axis points to the west. The Az angle is counted counterclockwise
  from the positive $x$-axis and refers to the direction of the primary
  $\gamma$-ray (figure \ref{figure1} (b)). Furthermore, the so-called US standard atmosphere
  is used as a model for the Earth's atmosphere.
\item[2.] The binary output of CORSIKA, containing information on the
  Cherenkov photon direction and its position on ground, is processed with a dedicated \textit{Reflector}
  program, which does the ray-tracing of the Cherenkov
  photons. To be able to adapt to different conditions without being forced to
  rerun CORSIKA, atmospheric absorption and scattering of Cherenkov photons as
  well as mirror condition is taken into account at this stage.
\item[3.] Finally, the output of the \textit{Reflector} program is
  processed by the \textit{Camera} program simulating the entire readout chain, i.e.
  PMT response, trigger and data acquisition system including electronic
  noise. Normally, a compact next-neighbour coincidence is required, i.e. at least
  four neighbouring pixels are required to trigger, and, if any of the pixels
  is taken out of the group, the remaining pixels are still neighbours.
  To adapt the MC data to the optical performance of the telescope,
  the simulation of the optical point spread function (PSF) can be tuned at this stage.
  The calibration and the image parameter calculation (Hillas analysis
  \cite{hil85}) is done using the MAGIC Analysis and Reconstruction Software
  (MARS) \cite{wag03}.
\end{itemize}

\subsection{Dedicated MC Production}\label{subsec:mcprod}

For the present studies only $\gamma$-rays were simulated. The MC data were produced following for
most instances the standard MC production of the MAGIC telescope as described beforehand.
All events were simulated as originating from a point source. By definition, the
telescope optical axis is always parallel to the direction of the primary
$\gamma$-ray. The impact parameter (IP) is defined as the distance from the
centre of the telescope mirror to the shower axis, which has the same
direction as the primary $\gamma$-ray.

\begin{figure}[!h]\centering
  \subfigure[]{\includegraphics[scale=.37]{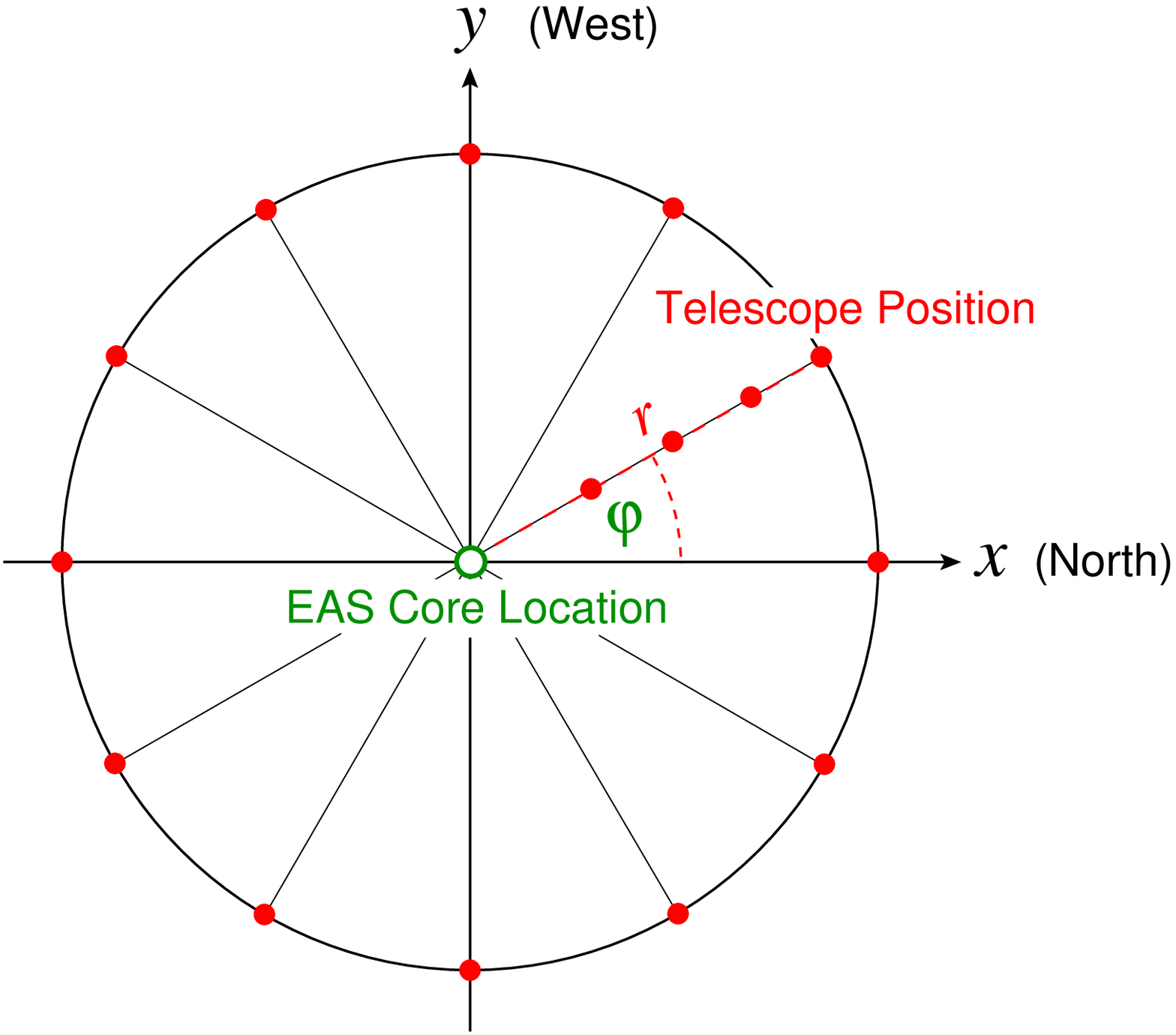}}\qquad
  \subfigure[]{\includegraphics[scale=.36]{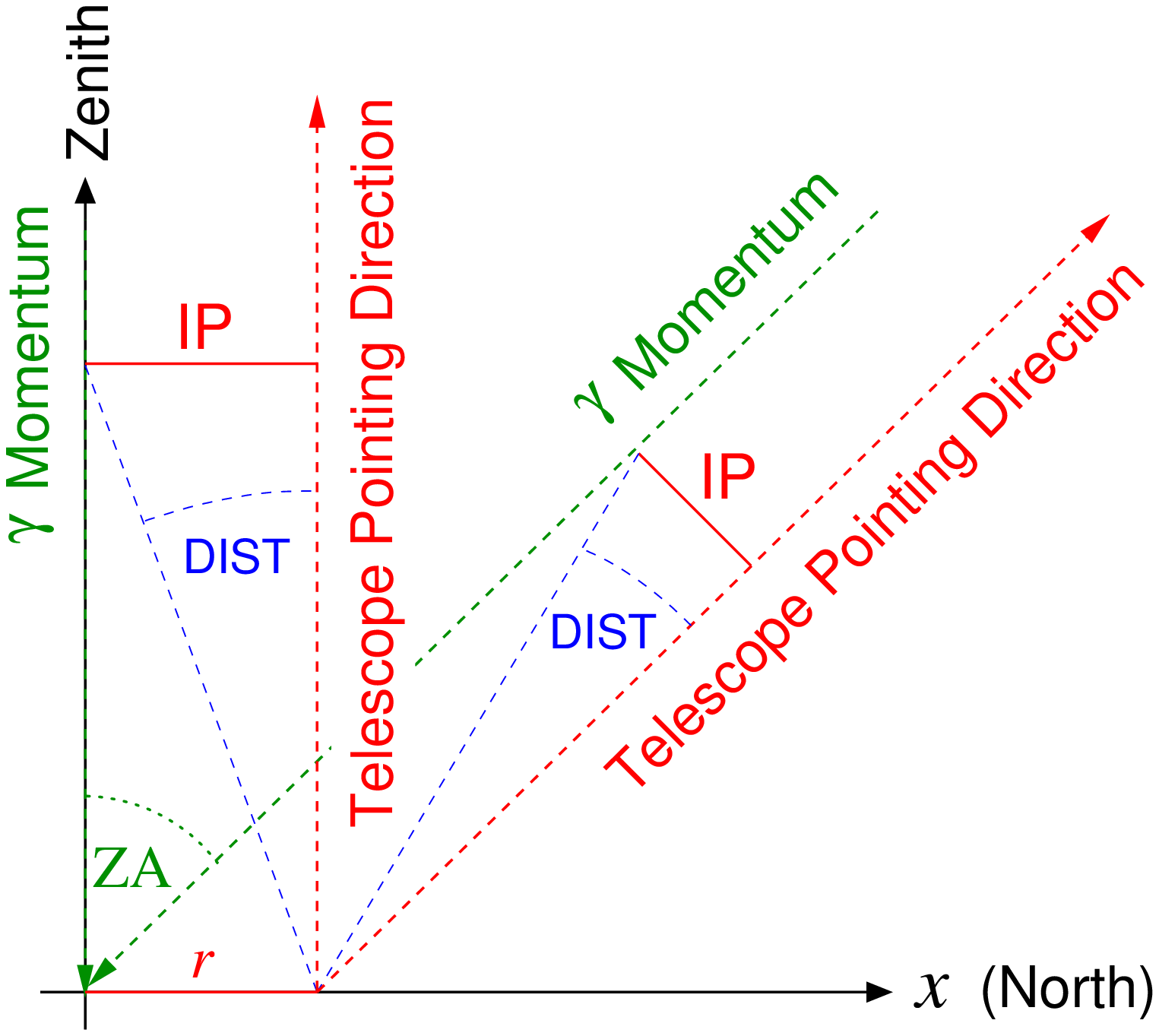}}
  \caption{(a) An illustration of the setup used in the MC simulations: the
    telescope positions are indicated as full circles and the EAS core location
    (the primary $\gamma$-ray impact point on the ground) as an open circle. The
    impact distance on ground $r$ is defined as the distance between the EAS core
    location and the telescope position. The telescope is situated always
    somewhere in the circle within which Cherenkov photons are kept.
    The simulations were done for fixed impact distances on ground $r=
    20$\,m,\dots 180\,m and angles $\varphi = 0^{\circ}$, $30^{\circ}$,\dots
    $330^{\circ}$. (b) The relation between the impact parameter (IP) and
    the impact distance $r$ on the ground is illustrated. In direction of the
    telescope's inclination the impact parameter equals $r\cos(\text{ZA})$.}
  \label{figure2}
\end{figure}

In contrast to the production of standard MC data, where the EAS core location is randomly placed
somewhere in a circle on the plane perpendicular to the direction of the EAS
(to estimate the telescope effective collection area), the EAS for this study were simulated
for fixed core locations and all Cherenkov photons arriving in a circle of 200\,m
radius were kept. This procedure reduces computing time because each CORSIKA
event can be used multiple times by placing the telescope (at the level of the
\textit{Reflector} program) somewhere into the
circle within which Cherenkov photons are kept (figure
\ref{figure2} (a)).
Besides, this approach allows to study the influence of the GF on the shower images
in great detail. For technical reasons the telescope was placed on equidistant
points concentrically to the EAS core location. By the choice of this setup, the impact parameter
varies like $r\sqrt{\cos^2(\text{Az}-\varphi)(\cos^2(\text{ZA})-1)+1}$
(figure \ref{figure2} (b)),
where $r$ is the distance between the EAS core location and the telescope
position.\\
The energy of the primary $\gamma$-ray was for different samples set to 30\,GeV, 50\,GeV, 70\,GeV,
120\,GeV, 170\,GeV, 300\,GeV, 450\,GeV or 1\,TeV, respectively.
The ZA was varied between $0^\circ$ and $60^\circ$ in steps of $20^\circ$, and
the Az angle between $0^\circ$ and $180^\circ$ in steps of
$30^\circ$, as the absolute value of the GF component normal to the EAS direction is
symmetric in the Az angle (figure \ref{figure1} (a)).
Hence, the maximum value for the angle $\theta$
achieved with the simulated telescope orientations is $87^\circ$.
The choice of discrete values for the $\gamma$-ray energy allows to
investigate the energy dispersion of the showers due to the GF.
The distance $r$ between the telescope position and the EAS core location
(impact point of the primary $\gamma$-ray on the ground) was varied between 20\,m and 180\,m in steps
of 20\,m and the angle $\varphi$ (figure \ref{figure2} (a)) between
$0^{\circ}$ and $330^{\circ}$ in steps of $30^{\circ}$, resulting in 108 configurations.\\
About $10^5$ events were simulated for each $\gamma$-ray energy, ZA and Az angle. As
a reference, MC data were also produced without GF. To be as realistic as
possible the MC simulations include the effects of photons from
the diffuse night sky background of $1.75\cdot
10^{12}\,\text{ph}\,\text{m}^{-2}\,\text{s}^{-1}\,\text{sr}^{-1}$
at the MAGIC site \cite{mir94} as well as electronic noise.

\section{Image Analysis}\label{sec:analysis}

\begin{figure}[!h] \centering
  \subfigure[]{\includegraphics[scale=.32]{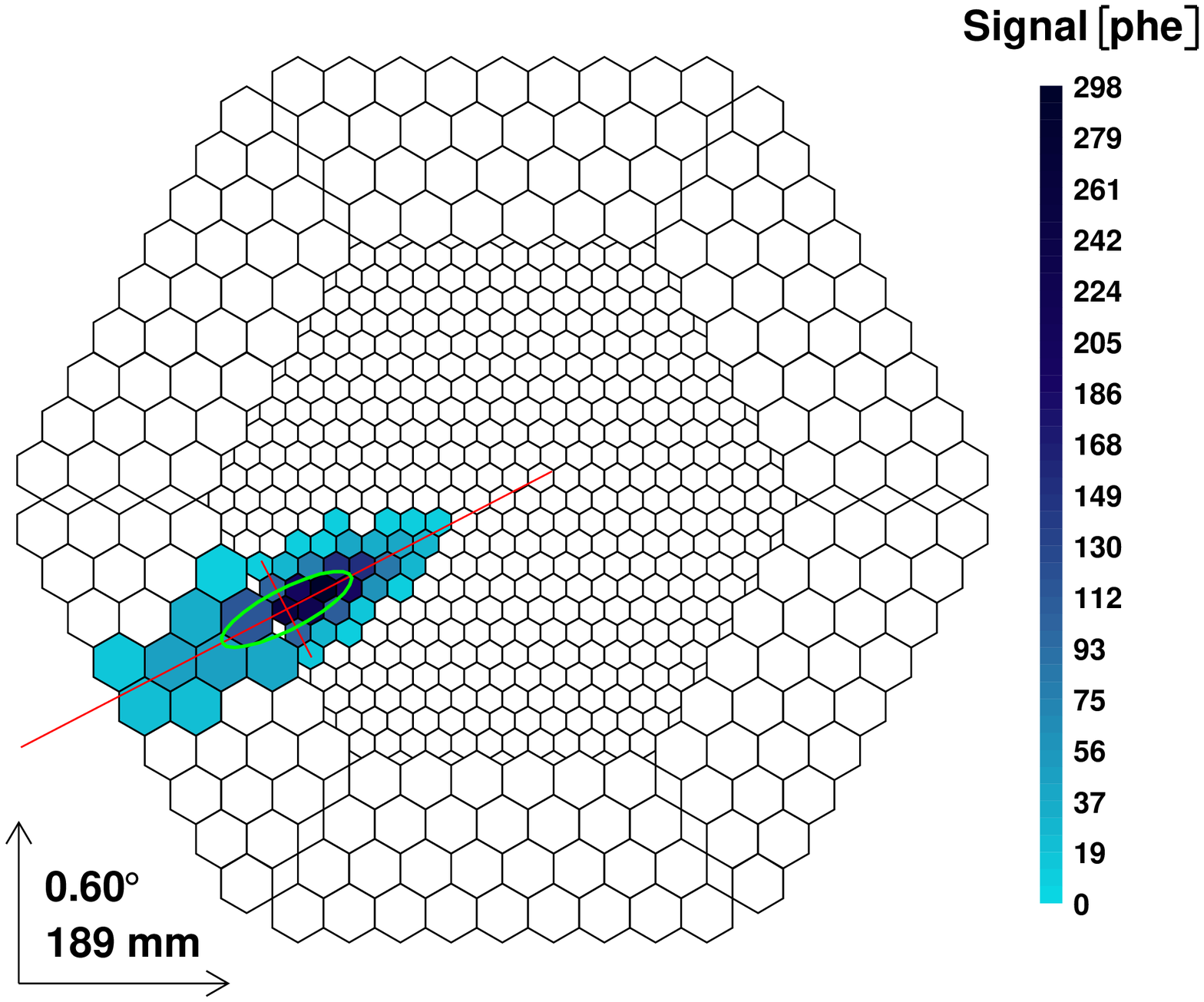}}\qquad
  \subfigure[]{\includegraphics[scale=.35]{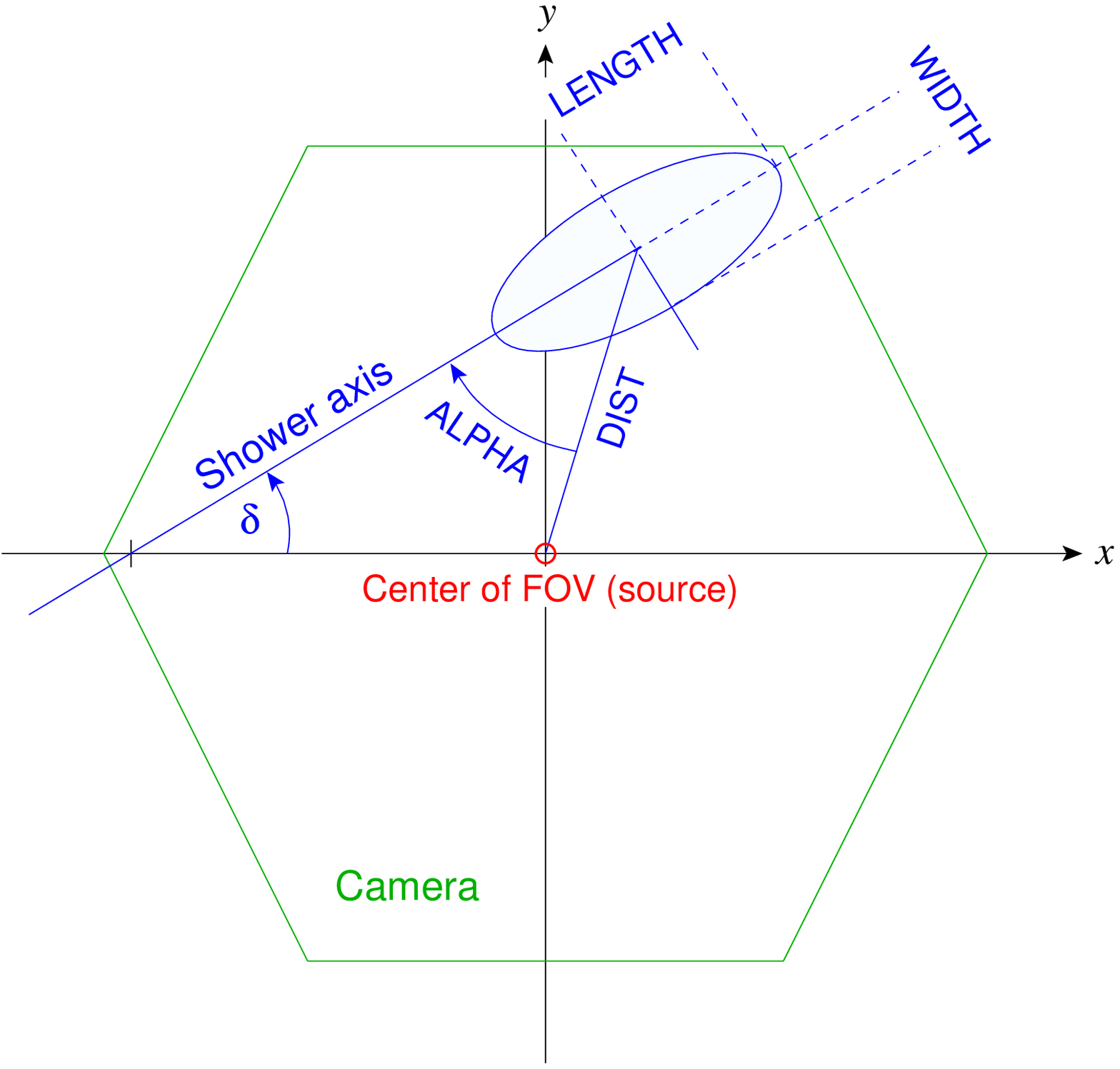}}
  \caption{(a) A MC-generated $\gamma$-ray shower image after application of the
  image cleaning. Only pixels surviving the image cleaning are used for the
  following steps of the analysis. (b) Definition of the image parameters. The
  light distribution is described by an ellipse whose major and minor axes
  describe the longitudinal and lateral spread of the Cherenkov light distribution.}
  \label{figure3}
\end{figure}

The MC-generated $\gamma$-ray showers were analysed using the standard MAGIC
software MARS \cite{wag03}.
Before parameterisation of the shower images a
tail-cut image cleaning was applied (figure \ref{figure3} (a)) \cite{alb0701}. 
The image cleaning requires the signals to be above a certain level.
For the MC studies presented here the
minimum required pixel content was 7\,photoelectrons (phe) for so-called core pixels and 4\,phe
for boundary pixels.\\
Shower images from $\gamma$-ray showers processed with the image cleaning are narrow and point towards the
source position in the field of view. To a first approximation the shower images
are elliptical and can be described by so-called Hillas parameters \cite{hil85}.
Detailed reviews on the imaging technique can be found elsewhere \cite{lan94,feg97}.
Some of the image parameters used for these studies are illustrated in figure \ref{figure3} (b).
The parameters WIDTH and LENGTH characterise the lateral and
longitudinal spread of the shower images (minor and major axes of the so-called
Hillas ellipse). Both parameters are very important since they allow a
powerful discrimination of $\gamma$-ray images against hadron induced
images.
The parameters DIST and ALPHA are related to the position and orientation of
shower images in the camera. The parameter DIST is directly related to the impact parameter of the primary
$\gamma$-ray (figure \ref{figure2} (b)).
The image parameter ALPHA is commonly used by standalone IACTs
to extract the $\gamma$-ray signal. ALPHA denotes the angle between the major
axis of the shower image and the vector connecting its centre of gravity with
the source position in the camera plane (camera centre).
It provides a very powerful discrimination
between $\gamma$-ray images from a point-like source and any isotropic background (mainly due to
hadrons), i.e. orientation discrimination.
The $\gamma$-ray signal from a VHE $\gamma$-ray source
under study appears as an excess at small values in the ALPHA parameter distribution.

Another viable image parameter is the so-called SIZE, which corresponds to the total
integrated light of a shower image after treatment with the image cleaning
procedure. It is therefore an estimate for the primary $\gamma$-ray energy.\\
The so-called DISP method \cite{fom94,les01} allows to reconstruct the arrival
direction of $\gamma$-ray candidates making use of the shape of
a shower image. The DISP parameter is determined according to the formula

\begin{equation}
  \text{DISP} = c_1(\text{SIZE}) + c_2(\text{SIZE})\cdot\frac{\text{WIDTH}}{\text{LENGTH}}.\label{eq:disp}
\end{equation}

Therein $c_1$ and $c_2$ are second-order polynomials optimised on MC simulated
$\gamma$-ray showers \cite{dom05}:

\begin{eqnarray*}
  c_1&=& +1.163^\circ + 0.542^\circ\left(\log_{10}(\text{SIZE})-2\right) - 0.672^\circ\left(\log_{10}(\text{SIZE})-2\right)^2,\\
  c_2&=& -0.265^\circ -  2.905^\circ\left(\log_{10}(\text{SIZE})-2\right) + 2.220^\circ\left(\log_{10}(\text{SIZE})-2\right)^2.
\end{eqnarray*}

In case of a single telescope the DISP method provides two possible solutions for the source position.
To overcome this ambiguity the asymmetry of the shower image
along the major image axis, which is related to the longitudinal development
of an EAS in the atmosphere, is used to reconstruct the true source position.
However, owing to false head-tail assignment the
percentage of correctly reconstructed events is typically limited to $\sim
80\,\%$, depending on the $\gamma$-ray energy \cite{dom05}. The outcome of the DISP analysis is usually
displayed in terms of a sky map of arrival directions. Because the DISP
parameter depends on both the eccentricity $\text{WIDTH}/\text{LENGTH}$ and on the orientation
of the shower images the influence of the GF on the shower development is
expected to degrade also the DISP method.\\
In this paper, we restrict ourselves to selected but representative results from the MC study.

\section{Results and Discussion}

\subsection{GF Effects on Shape and Orientation of  $\gamma$-ray Shower Images}

\begin{figure}[!h] 
  \subfigure[$\text{Az}=0^\circ$, $\theta=12^\circ$.]{
    \includegraphics[scale=.34]{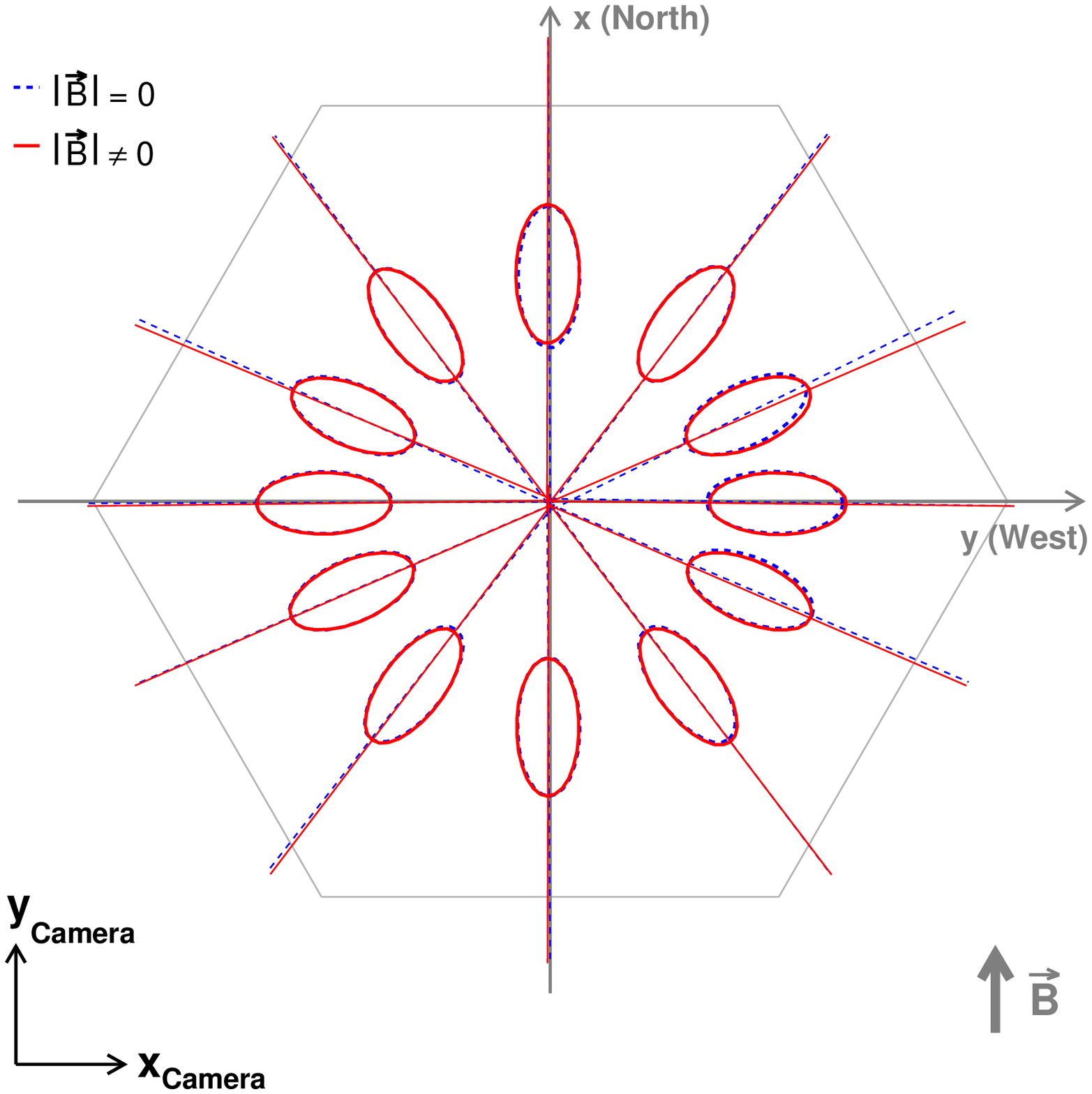}}\quad
  \subfigure[$\text{Az}=180^\circ$, $\theta=87^\circ$.]{
    \includegraphics[scale=.34]{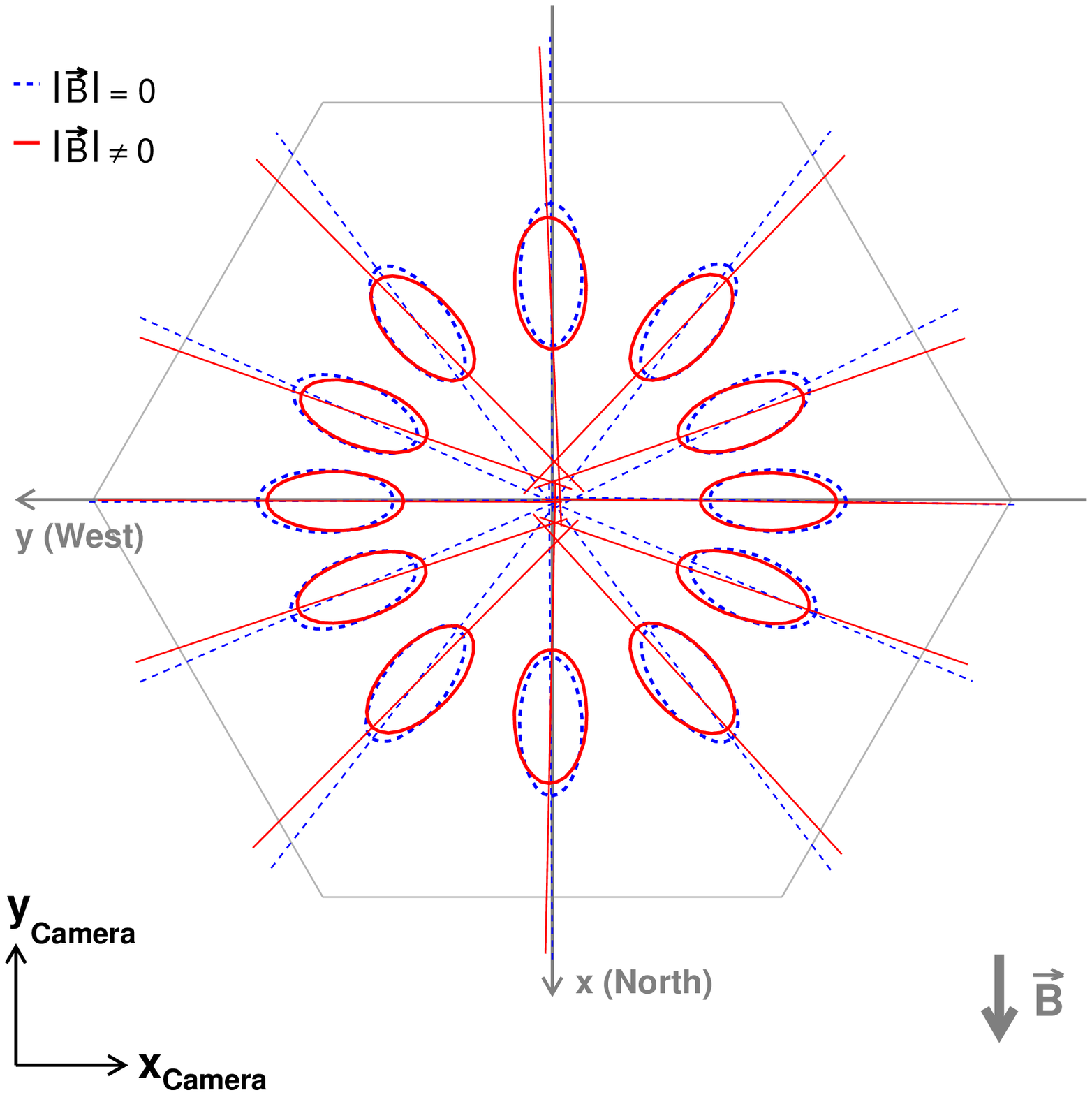}}
  \caption{Shower images in the telescope camera for 450\,GeV $\gamma$-rays,
    100\,m impact parameter, $\text{ZA}=40^\circ$, $\text{Az}=0^\circ$ and $180^\circ$, respectively.
    The images indicated by solid lines were obtained
    for enabled GF and the ones indicated by dashed lines were obtained for disabled GF.
    The orientation of the GF component normal to
    the direction of the EAS (telescope pointing direction) is indicated in
    the lower right part of the figures (see text for more details).}
  \label{figure4}
\end{figure}

The orientation of the $\gamma$-ray images from a source under study is
used for the suppression of the isotropic hadronic background. It is accordingly
important to study how the GF influences the orientation of shower images.
Figure \ref{figure4} shows shower images (Hillas ellipses)
in the telescope camera for 450\,GeV $\gamma$-rays, 100\,m impact parameter,
$\text{ZA}=40^\circ$, and different Az angles between $0^\circ$
and $180^\circ$. The ellipses drawn with solid lines were obtained for enabled GF
and the ones drawn with dashed lines for disabled GF in the MC simulation.
For each angle $\varphi = 0^{\circ}$, $30^{\circ}$,\dots $330^{\circ}$ the
size, the position and the orientation of the ellipse in the camera was
determined by taking the mean values from the corresponding Hillas parameter
distribution.
The ellipses are superimposed on the projection of the CORSIKA coordinate
system, whose $x$-axis is aligned with the magnetic north.
The direction of the GF component normal to the direction of the EAS is indicated in
the lower right part of the figures.

\begin{figure}[!h] \centering
  \subfigure[$\text{Az}=0^\circ$, $\text{ZA}=0^\circ$, $\text{IP}= 40\,\text{m}$, $\theta = 52^\circ$.]{
    \includegraphics[scale=.31]{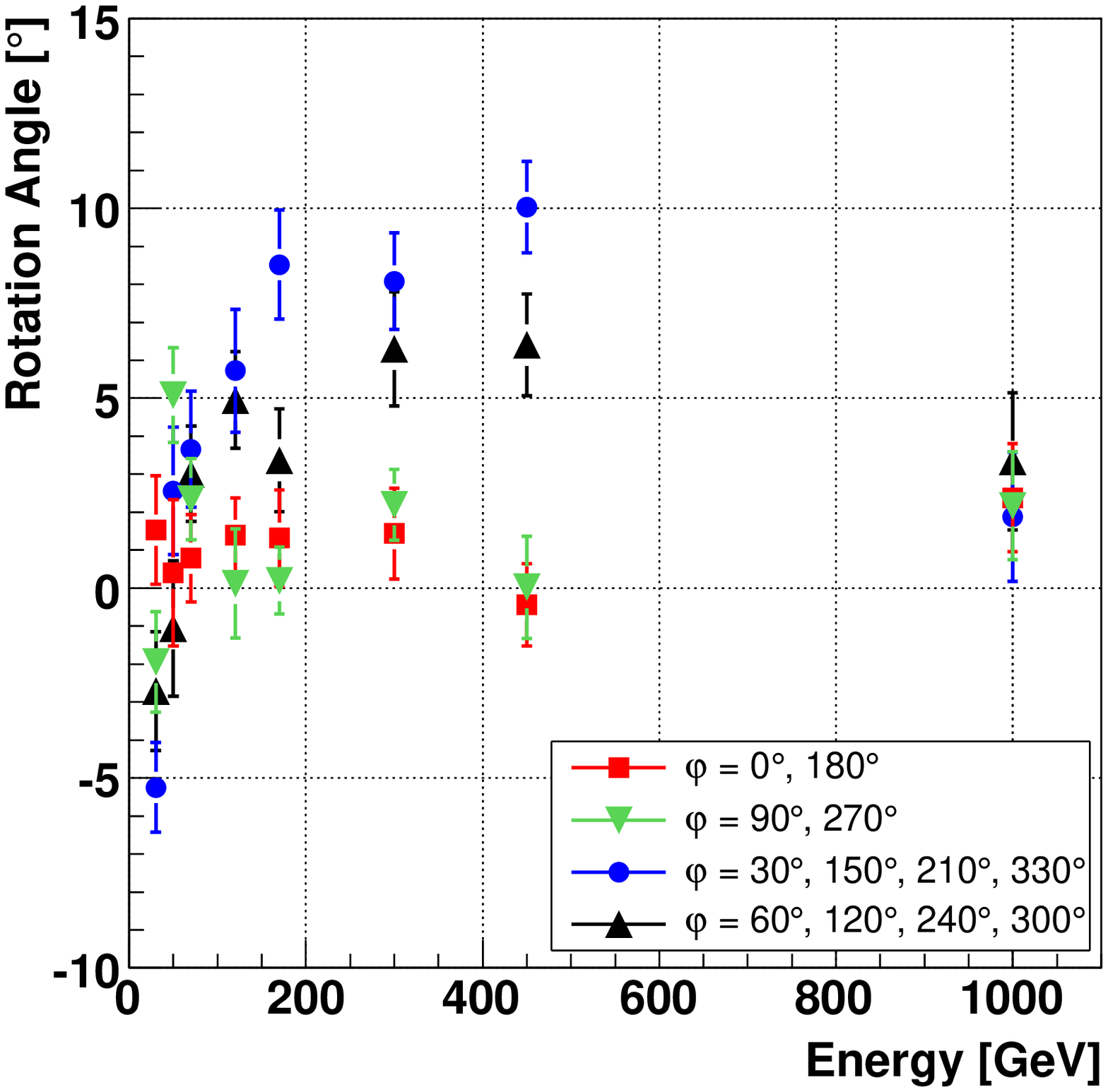}}\quad
  \subfigure[$\text{Az}=0^\circ$, $\text{ZA}=0^\circ$, $\text{IP}= 120\,\text{m}$, $\theta = 52^\circ$.]{
    \includegraphics[scale=.31]{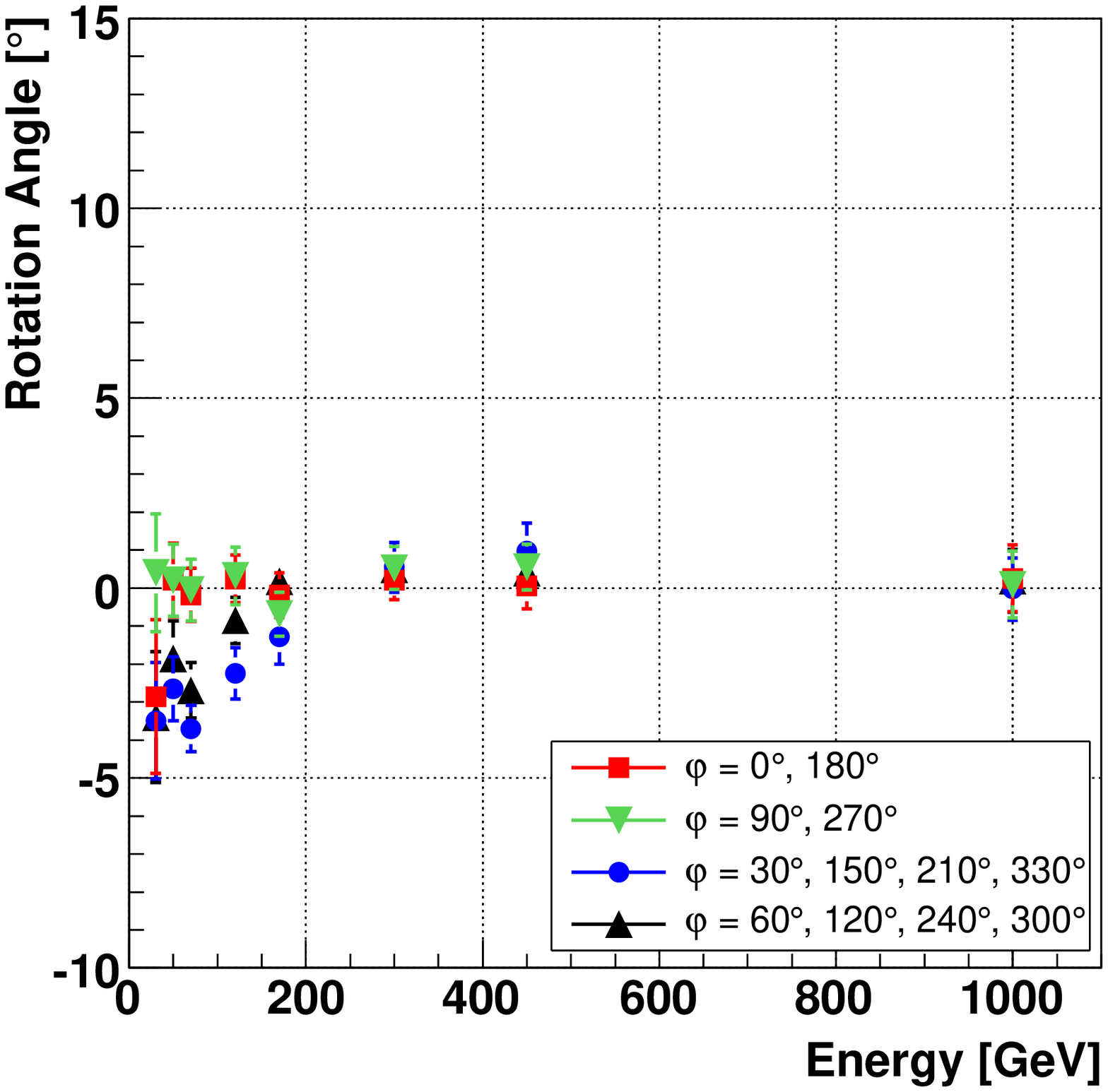}}\\
  \subfigure[$\text{Az}=180^\circ$, $\text{ZA}=40^\circ$, $\text{IP}\approx 40\,\text{m}$, $\theta = 87^\circ$.]{
    \includegraphics[scale=.31]{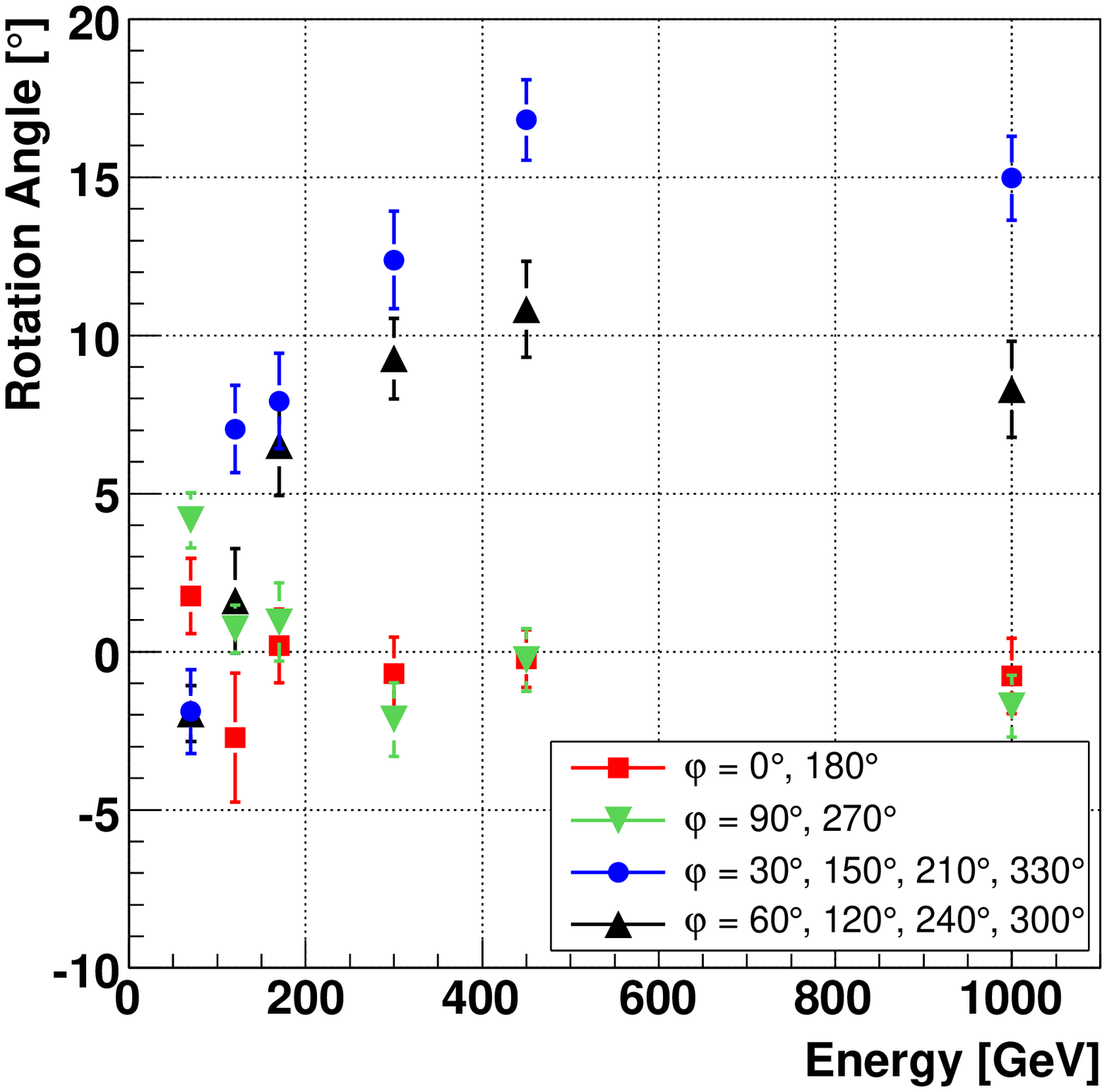}}\quad
  \subfigure[$\text{Az}=180^\circ$, $\text{ZA}=40^\circ$, $\text{IP}\approx
  120\,\text{m}$, $\theta = 87^\circ$.]{
    \includegraphics[scale=.31]{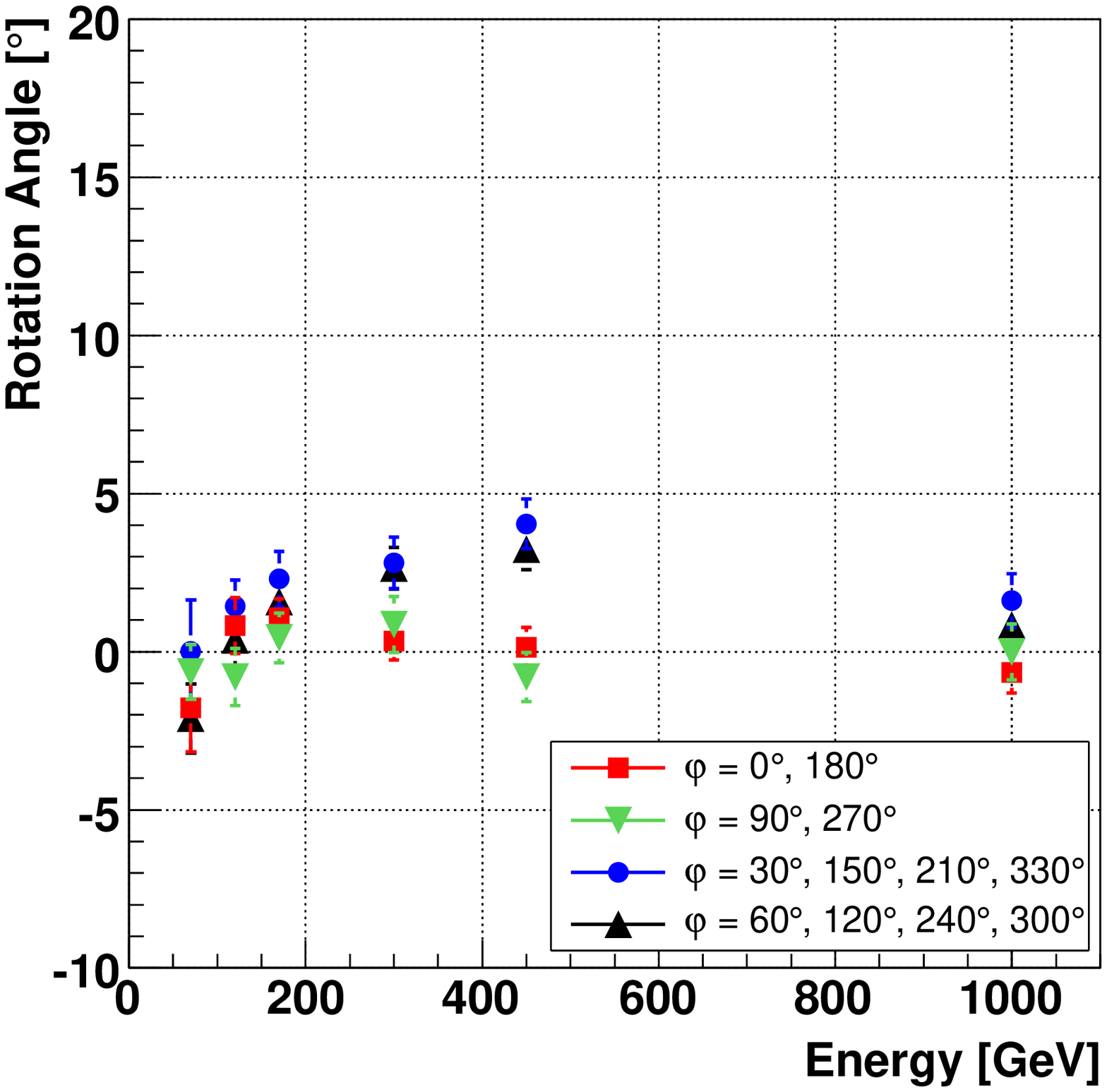}}
  \caption{The rotation angle of $\gamma$-ray images versus energy for 40\,m
    and 120\,m impact parameter, $\text{ZA}=0^\circ$, $40^\circ$,
    $\text{Az}=0^\circ$ and $180^\circ$. Full triangle down data points correspond to the case
    where the connecting line between the EAS core location on ground and the
    telescope position is parallel to the north-south direction ($x$-axis),
    while for full square data points the telescope is situated on the
    $y$-axis. Full triangle up and full circle data points 
    correspond to intermediate telescope positions.}
  \label{figure5}
\end{figure}

\clearpage

\begin{figure}[!ht] \centering
  \subfigure[$\text{Az}=0^\circ$, $\text{ZA}=20^\circ$, 120\,GeV energy, $\theta = 32^\circ$.]{
    \includegraphics[scale=.31]{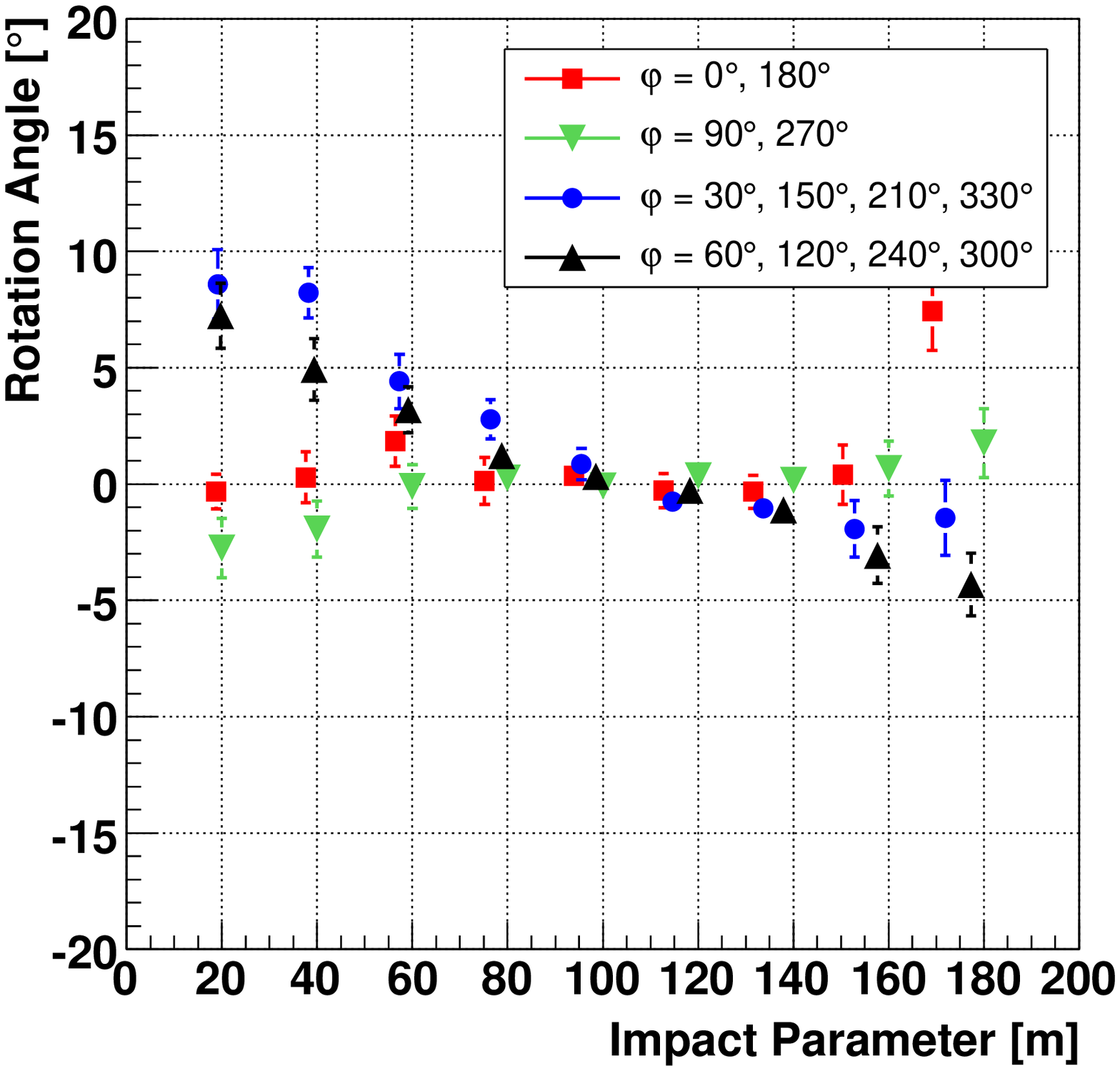}}\quad
  \subfigure[$\text{Az}=180^\circ$, $\text{ZA}=20^\circ$, 120\,GeV energy, $\theta = 72^\circ$.]{
    \includegraphics[scale=.31]{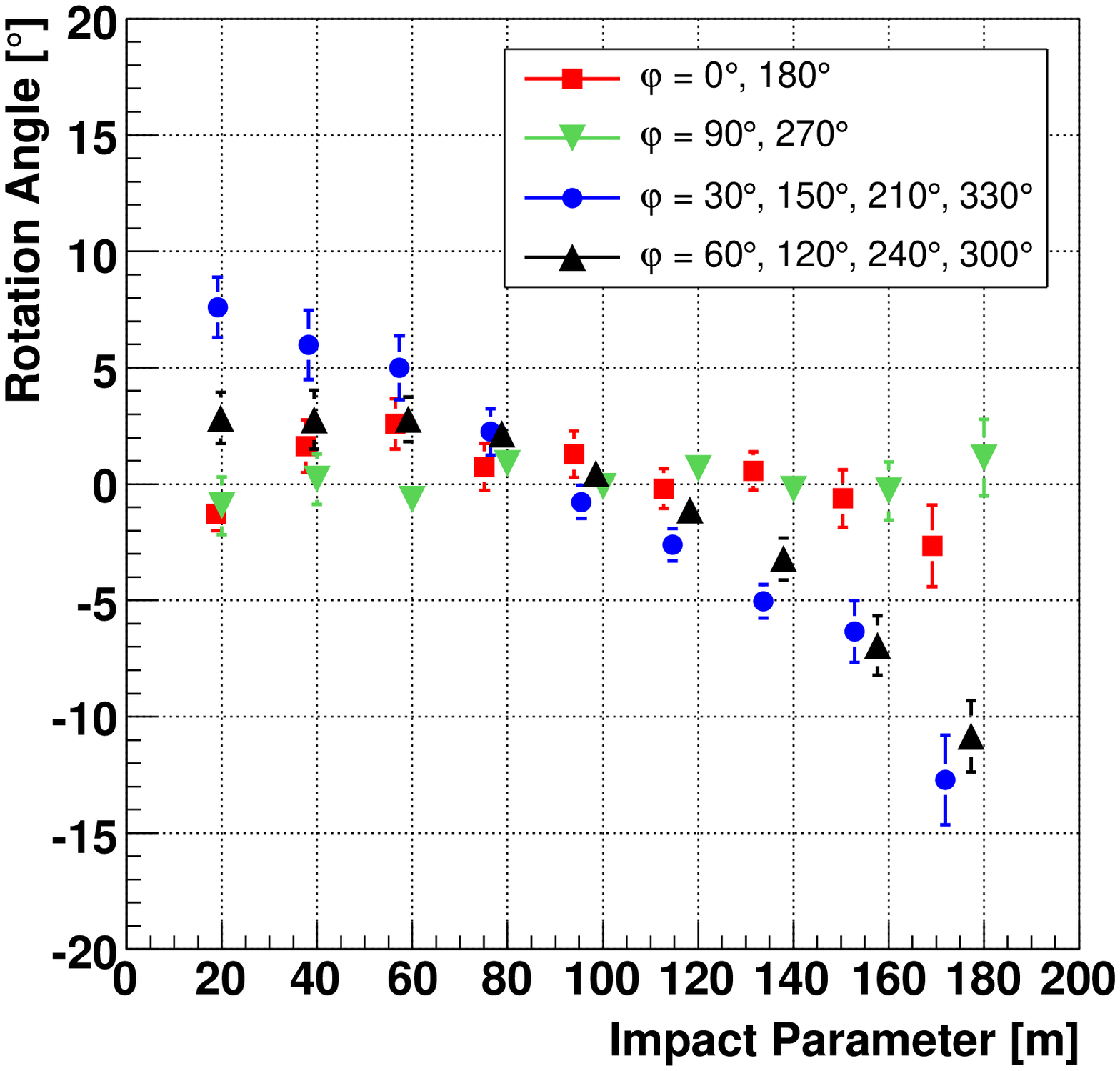}}\\
  \subfigure[$\text{Az}=0^\circ$, $\text{ZA}=40^\circ$, 450\,GeV energy, $\theta = 12^\circ$.]{
    \includegraphics[scale=.31]{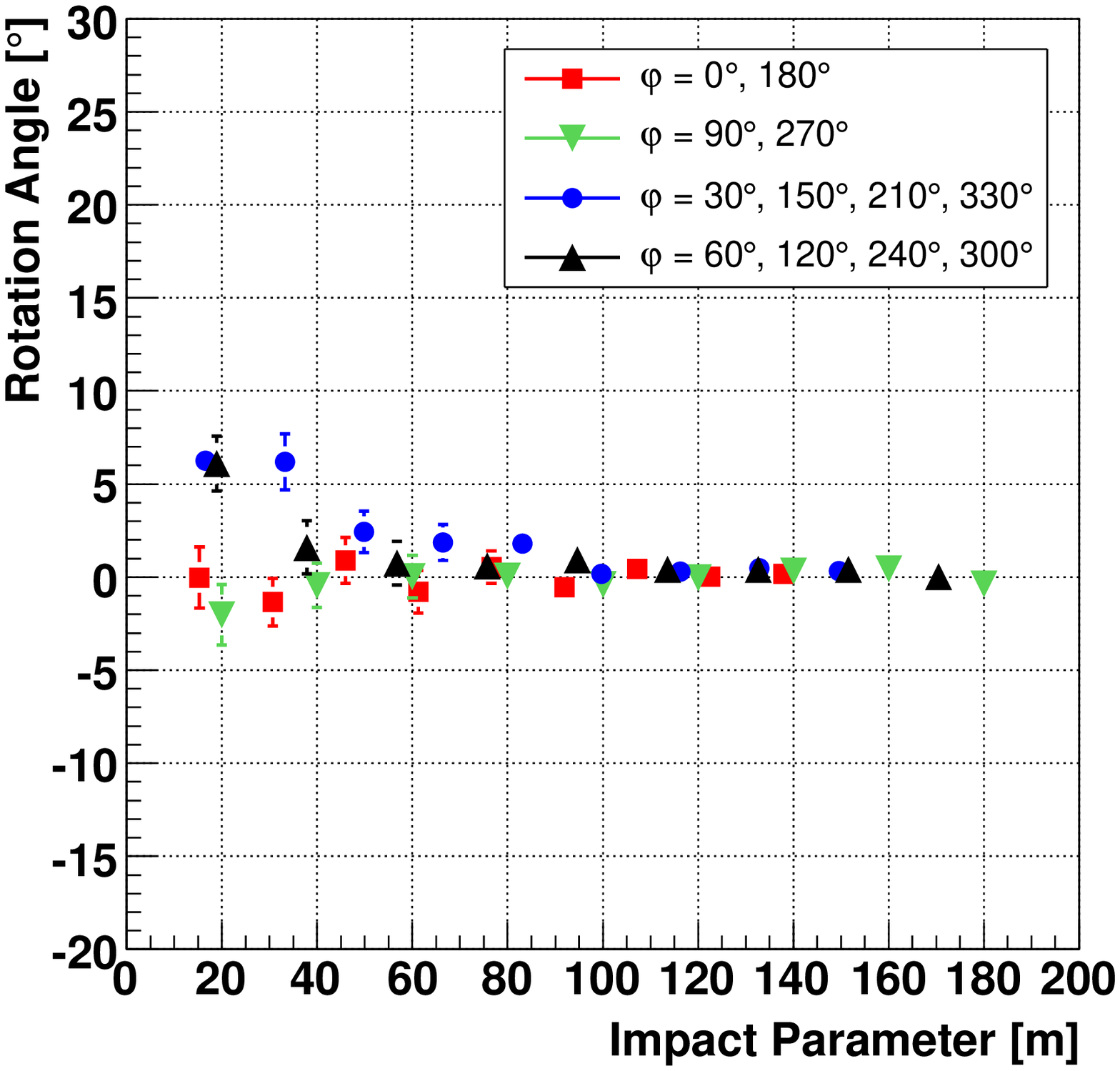}}\quad
  \subfigure[$\text{Az}=180^\circ$, $\text{ZA}=40^\circ$, 450\,GeV energy, $\theta = 87^\circ$.]{
    \includegraphics[scale=.31]{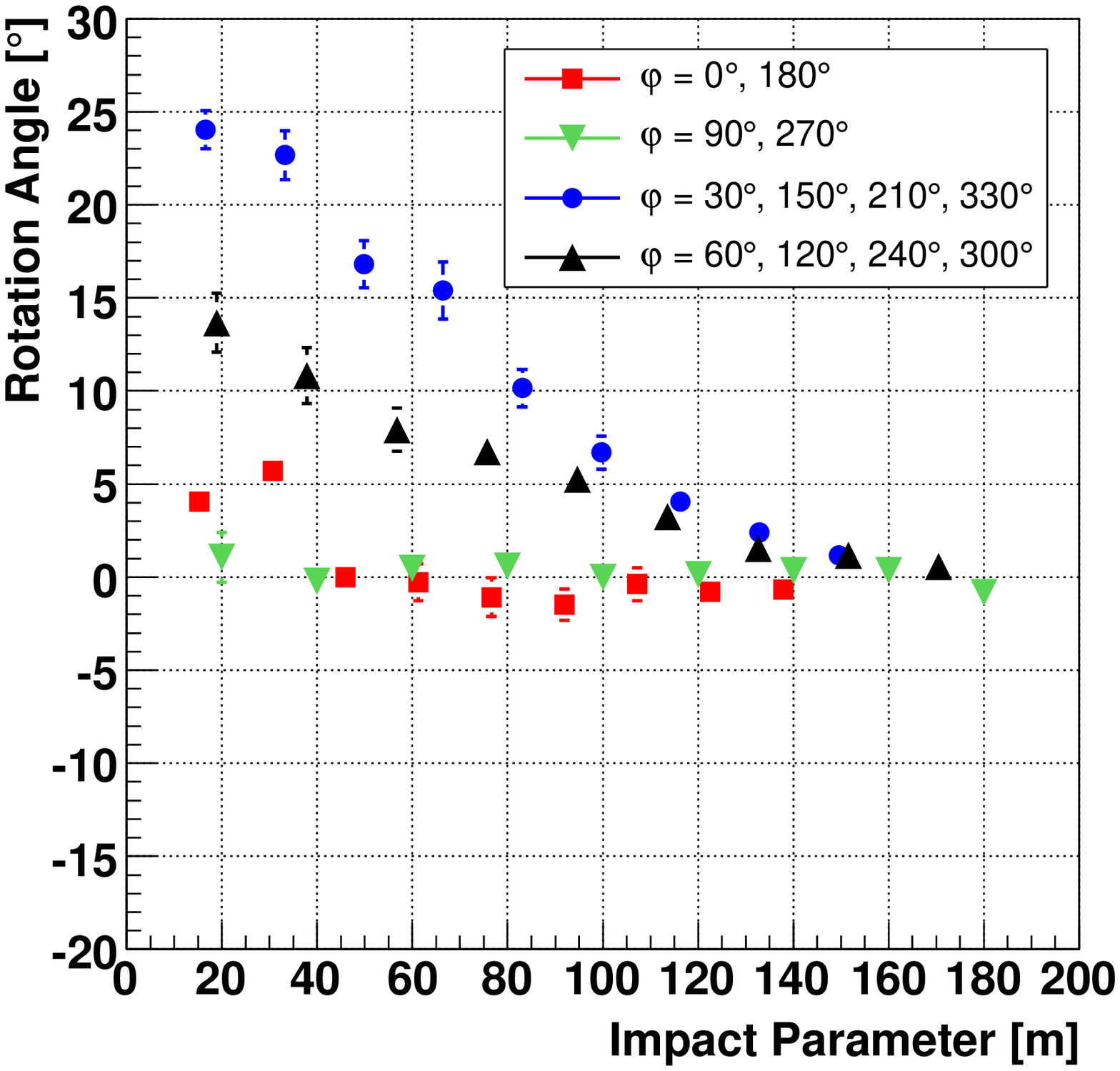}}\\
  \subfigure[$\text{Az}=0^\circ$, $\text{ZA}=40^\circ$, 1\,TeV energy, $\theta = 12^\circ$.]{
    \includegraphics[scale=.31]{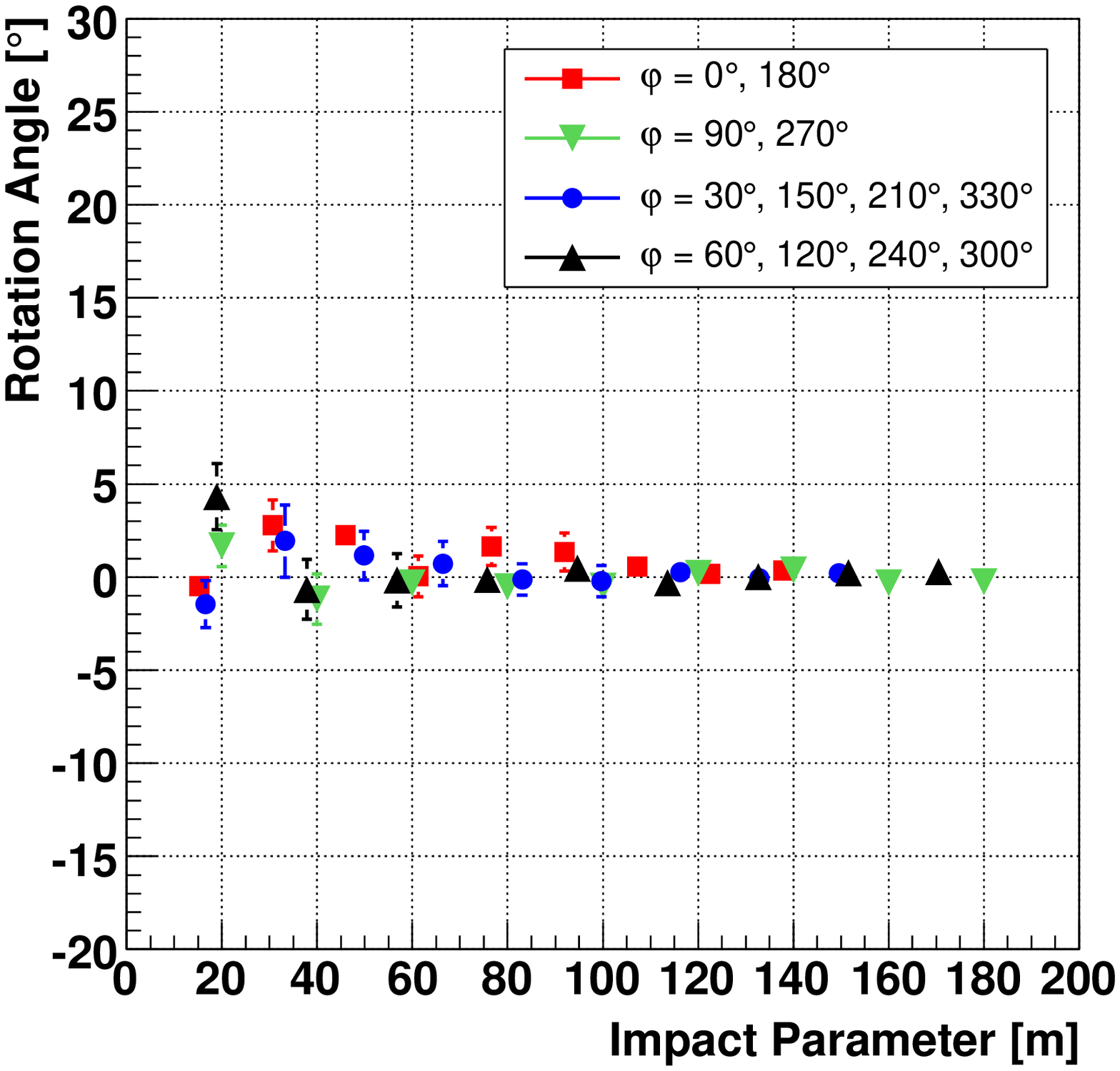}}\quad
  \subfigure[$\text{Az}=180^\circ$, $\text{ZA}=40^\circ$, 1\,TeV energy, $\theta = 87^\circ$.]{
    \includegraphics[scale=.31]{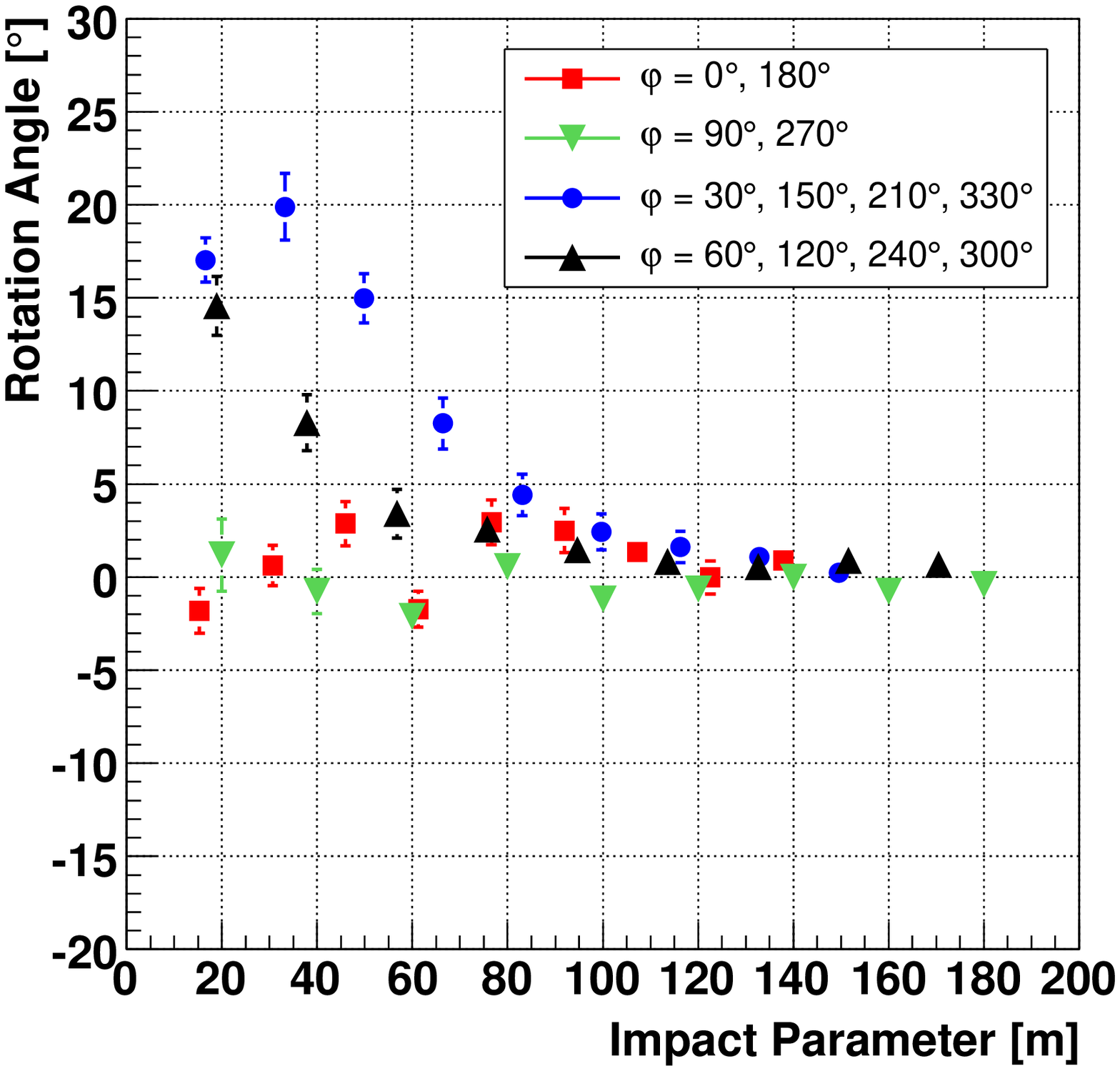}}
  \caption{The rotation angle of $\gamma$-ray images versus impact parameter
    for $\text{ZA}=20^\circ$ and $40^\circ$, $\text{Az}=0^\circ$ and $180^\circ$,
    120\,GeV, 450\,GeV and 1\,TeV energy (see text for more details).}
  \label{figure6}
\end{figure}

\clearpage

\begin{figure}[!t] \centering
  \subfigure[$\text{Az}=0^\circ$, $\theta=12^\circ$.]{
    \includegraphics[scale=.31]{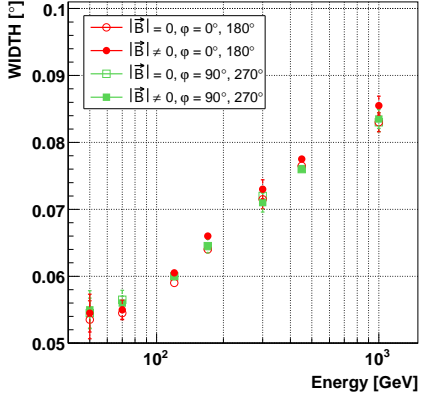}}\quad
  \subfigure[$\text{Az}=180^\circ$, $\theta=87^\circ$.]{
    \includegraphics[scale=.31]{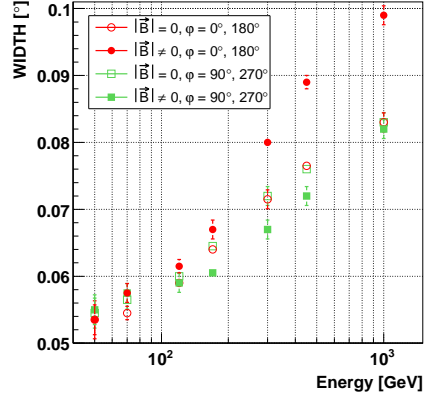}}\\
  \subfigure[$\text{Az}=0^\circ$, $\theta=12^\circ$.]{
    \includegraphics[scale=.31]{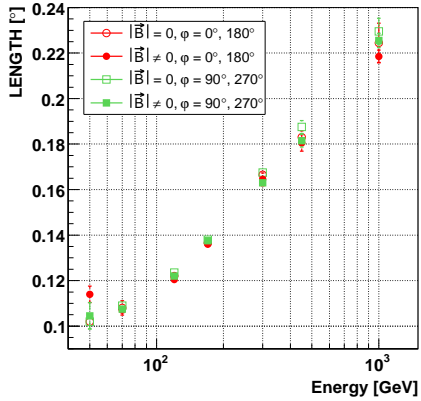}}\quad
  \subfigure[$\text{Az}=180^\circ$, $\theta=87^\circ$.]{
    \includegraphics[scale=.31]{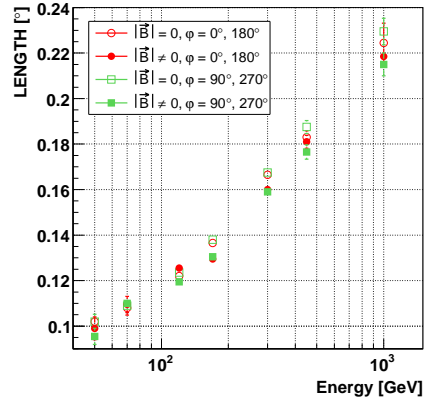}}
  \caption{The average WIDTH and LENGTH of $\gamma$-ray images versus energy
    for 120\,m impact parameter, $\text{ZA}=40^\circ$,
    $\text{Az}=0^\circ$ and $180^\circ$ (see text for more details).}
  \label{figure7}
\end{figure}

As can be seen the average orientation is preserved only for shower images oriented either
parallel or normal to the projected direction of the GF in the camera,
which is by definition the GF component normal to the direction of the EAS (telescope pointing direction).
Shower images situated at intermediate angles
are systematically rotated away from the projected direction of the GF.
The sideways spread of the images result in a systematic rotation away from
the camera centre (source position). This was also reported in \cite{cha00}.
The magnitude of the rotation depends not only on the angle $\theta$ between the axis of the
EAS and the direction of the GF but mainly on the core position of the EAS with
respect to the telescope, which is given by the angle $\varphi$.\\
By comparing the orientations for images generated with disabled GF to the
ones for enabled GF it is possible to determine the rotation angle.
Figure \ref{figure5} shows the rotation
angle of $\gamma$-ray images versus energy for an impact parameter of 40\,m and 120\,m.
The figures show that the rotation angle depends on the impact parameter and on the
$\gamma$-ray energy.
However, the average rotation angle for images oriented either parallel or normal
to the direction of the GF in the camera is zero (full square and full triangle down data points).
Images oriented at intermediate angles are systematically rotated.
For $\text{ZA}=0^\circ$, $\text{Az}=0^\circ$ and 40\,m impact parameter (figure \ref{figure5}
(a)) and for $\text{ZA}=40^\circ$, $\text{Az}=180^\circ$ (figure \ref{figure5}
(c) and (d)), which is the most unfavourable telescope pointing direction, the rotation angle
is maximal for $\gamma$-ray energies around 450\,GeV.\\
Figure \ref{figure6} illustrates the
dependency of the rotation angle on the impact parameter. Small impact
parameters correspond to images with low eccentricity
$\text{WIDTH}/\text{LENGTH}$, which may be rotated through a large
angle. Figure \ref{figure6} (b) also
shows that the direction of the rotation depends on the impact parameter.
The shower images are not always rotated away from the
direction of the GF in the camera but can even be rotated towards it.
Consequently, the correction of observational data for GF effects by de-rotating the shower images
must take into account the energy dependence of the rotation angle and its
dependence on the impact parameter (DIST). For the most unfavourable telescope pointing
direction and for small impact parameters even 1\,TeV shower images are rotated
through a large angle (figure
\ref{figure6} (f)).
The influence of the GF on the shape of the $\gamma$-ray shower images also depends
on the primary $\gamma$-ray energy, the impact parameter and the orientation of the EAS
relative to the direction of the GF.
Figure \ref{figure7} shows the
average WIDTH and LENGTH of $\gamma$-ray images versus energy for 120\,m
impact parameter, $\text{ZA}=40^\circ$, and $0^\circ$ as well as $\text{Az}=180^\circ$.
For the most unfavourable telescope pointing direction, i.e. $\text{ZA}=40^\circ$ and
$\text{Az}=180^\circ$ ($\theta = 87^\circ$) significant GF effects on the image
parameter WIDTH occur for $\gamma$-ray energies above $\sim 100\,\text{GeV}$
(figure \ref{figure7} (b)).
The influence of the GF on the image parameter WIDTH is
considerably larger than on LENGTH.
Due to the influence of the GF the average WIDTH is increased for images
where the connecting line between the EAS core location on ground and the
telescope position is parallel to the magnetic north-south direction (telescope situated on the
$x$-axis, full and open circle data points). Images aligned with the direction of the GF in the camera are
horizontally stretched compared to the situation of disabled GF in the MC,
whereas images oriented normal to the direction of the GF have a smaller WIDTH
and are thus elongated due to the influence of the GF (telescope situated on the
$y$-axis, full and open square data points).

\subsection{GF Effects on the Image Parameter ALPHA}

In the preceding section it was shown that the GF can strongly alter the
average shape and orientation of $\gamma$-ray shower images in the camera.
Even though the orientation of shower images and the image parameter ALPHA are
correlated it is important to investigate the influence of the GF on the image
parameter providing the most powerful discrimination between $\gamma$-rays
from a point-like source and unwanted isotropic background (mainly hadrons).

\begin{figure}[!h] \centering
  \subfigure[$\text{IP}=40\,\text{m}$.]{
    \includegraphics[scale=.33]{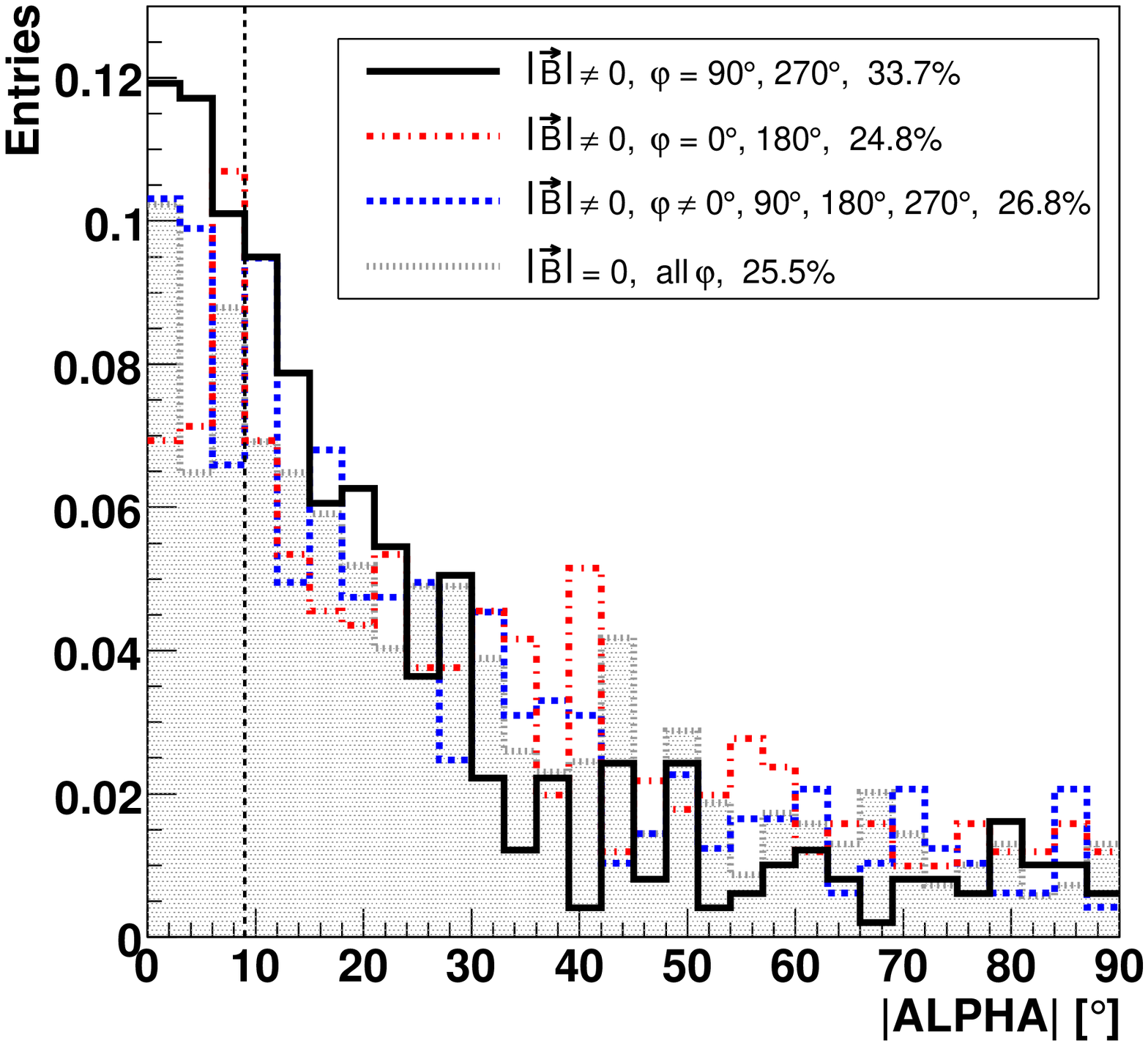}}\quad
  \subfigure[$\text{IP}=120\,\text{m}$.]{
    \includegraphics[scale=.33]{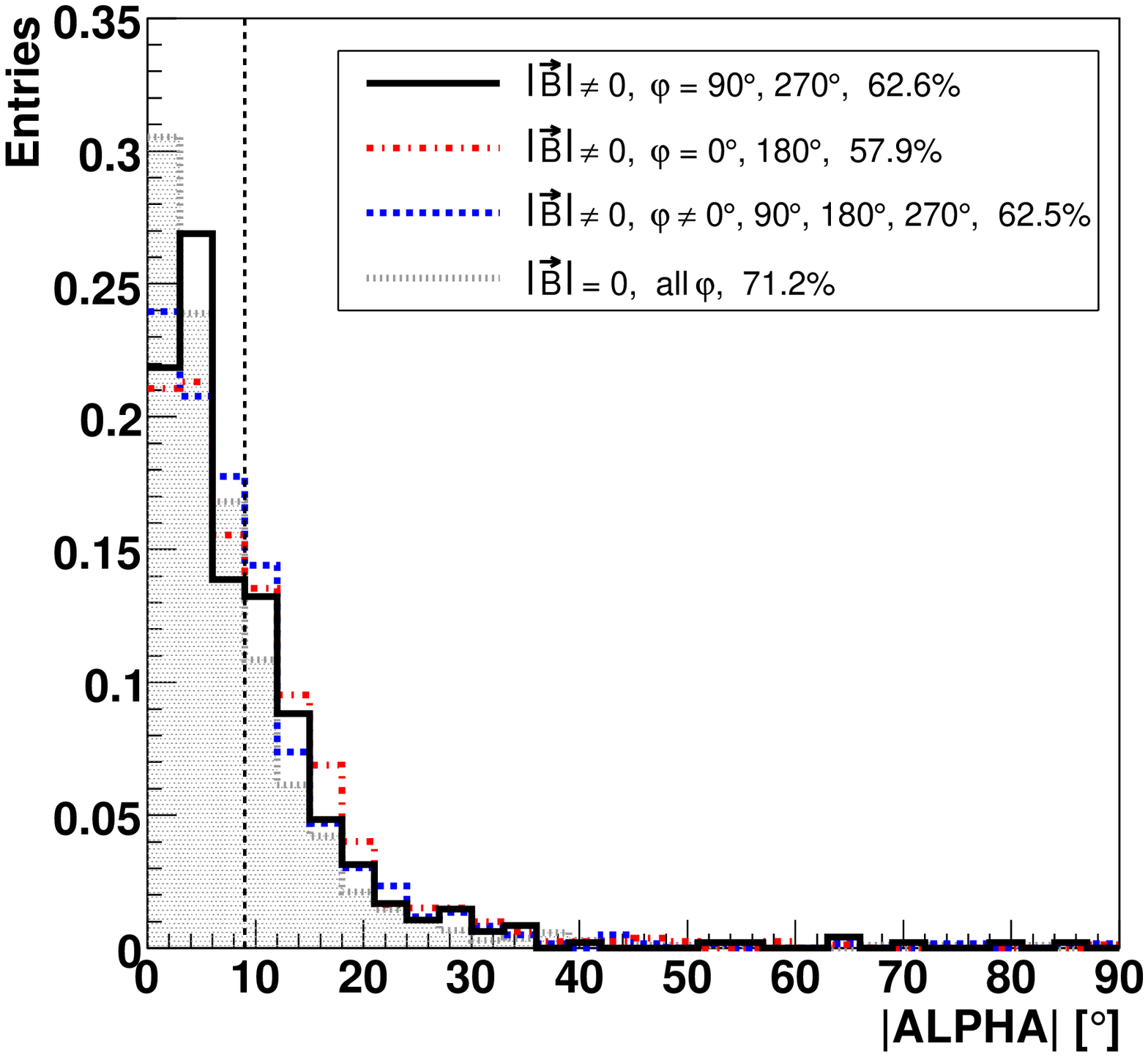}}
  \caption{Normalised distributions of the image parameter ALPHA for 50\,GeV
    $\gamma$-rays, $\text{ZA}=0^\circ$, $\text{Az}=0^\circ$ ($\theta=52^\circ$),
    40\,m (a) and 120\,m impact parameter (b). Different configurations are
    compared: the distributions indicated by solid lines correspond to
    $\varphi=90^\circ$ and $270^\circ$, the dash-dotted distributions to $\varphi=0^\circ$ and $180^\circ$, and
    the distributions indicated by dashed lines to intermediate telescope
    positions $\varphi = 30^\circ$, $60^\circ$, $120^\circ$, $150^\circ$, $210^\circ$,
    $240^\circ$, $300^\circ$ and $330^\circ$.
    The corresponding distribution obtained without GF in the MC simulation is also plotted
    (dotted line). The percentage of events with $|\text{ALPHA}|\leq 9^\circ$ is given in the legend.}
  \label{figure8}
\end{figure}

\begin{figure}[!h] \centering
  \subfigure[$\text{Az}=0^\circ$, $\text{IP}\approx 40\,\text{m}$, $\theta=12^\circ$.]{
    \includegraphics[scale=.33]{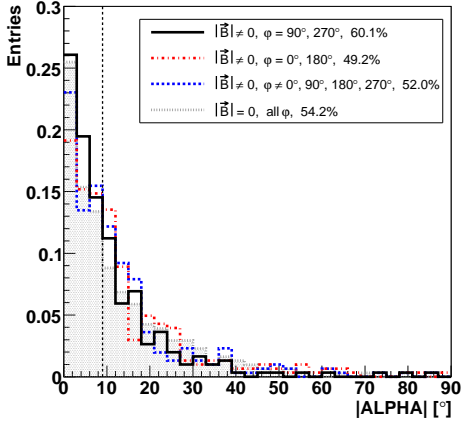}}\quad
  \subfigure[$\text{Az}=180^\circ$, $\text{IP}\approx 40\,\text{m}$, $\theta=87^\circ$.]{
    \includegraphics[scale=.33]{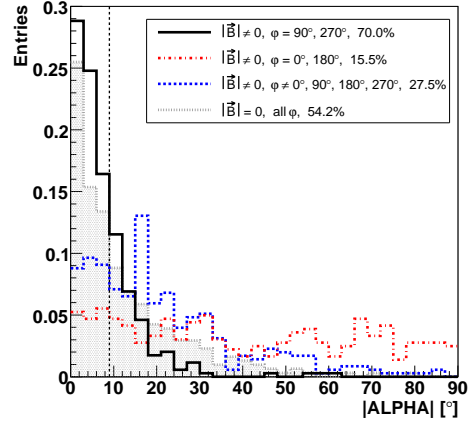}}\\
  \subfigure[$\text{Az}=0^\circ$, $\text{IP}\approx 120\,\text{m}$, $\theta=12^\circ$.]{
    \includegraphics[scale=.33]{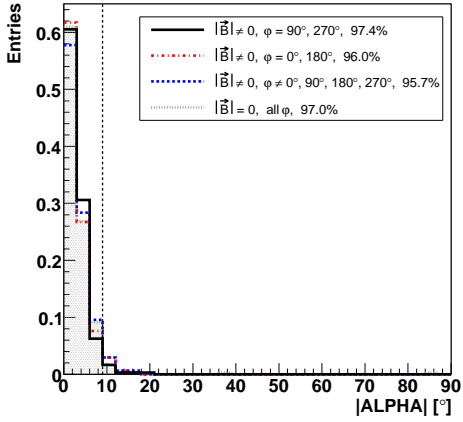}}\quad
  \subfigure[$\text{Az}=180^\circ$, $\text{IP}\approx 120\,\text{m}$, $\theta=87^\circ$.]{
    \includegraphics[scale=.33]{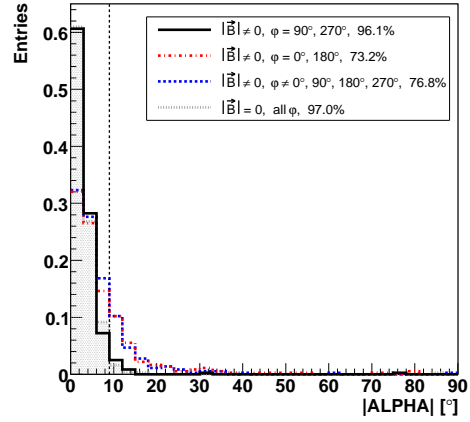}}
  \caption{As figure \ref{figure8}, but for 450\,GeV
    $\gamma$-rays, $\text{ZA}=40^\circ$, $\text{Az}=0^\circ$ and $180^\circ$.}
  \label{figure9}
\end{figure}

\begin{figure}[!h] \centering
  \subfigure[$\text{Az}=0^\circ$, $\text{IP}\approx 40\,\text{m}$, $\theta=12^\circ$.]{
    \includegraphics[scale=.33]{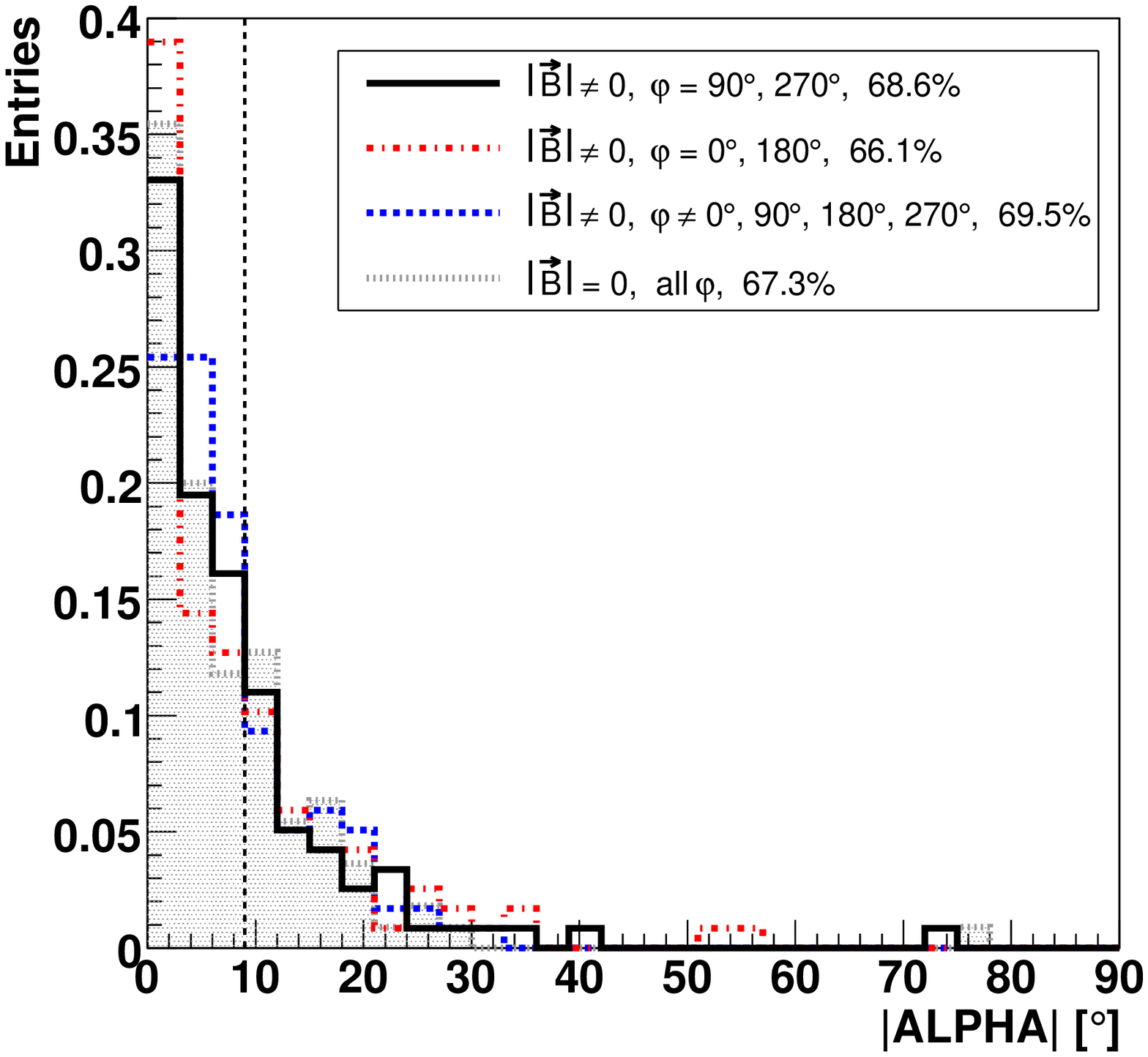}}\quad
  \subfigure[$\text{Az}=180^\circ$, $\text{IP}\approx 40\,\text{m}$, $\theta=87^\circ$.]{
    \includegraphics[scale=.33]{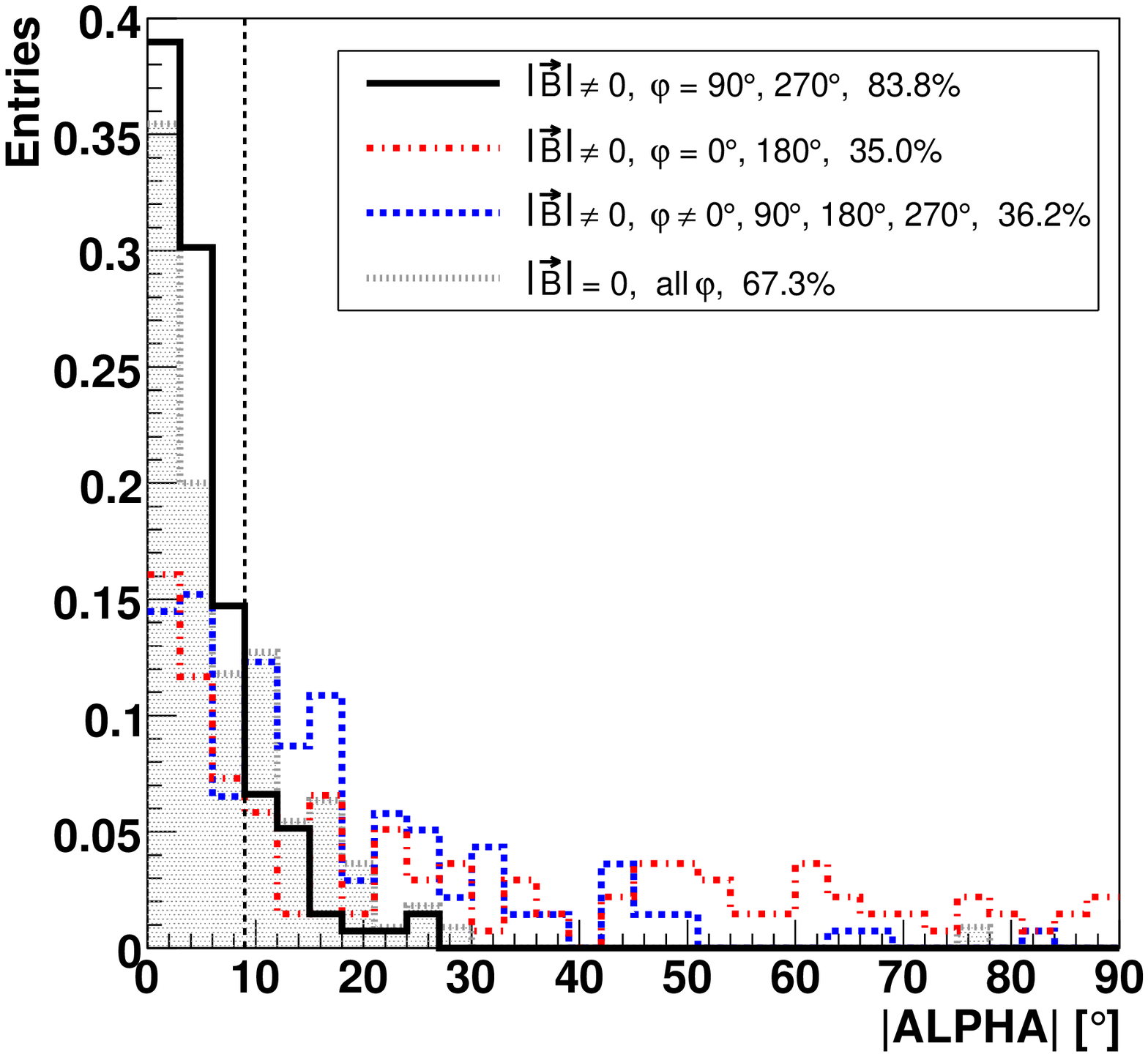}}\\
  \subfigure[$\text{Az}=0^\circ$, $\text{IP}\approx 120\,\text{m}$, $\theta=12^\circ$.]{
    \includegraphics[scale=.33]{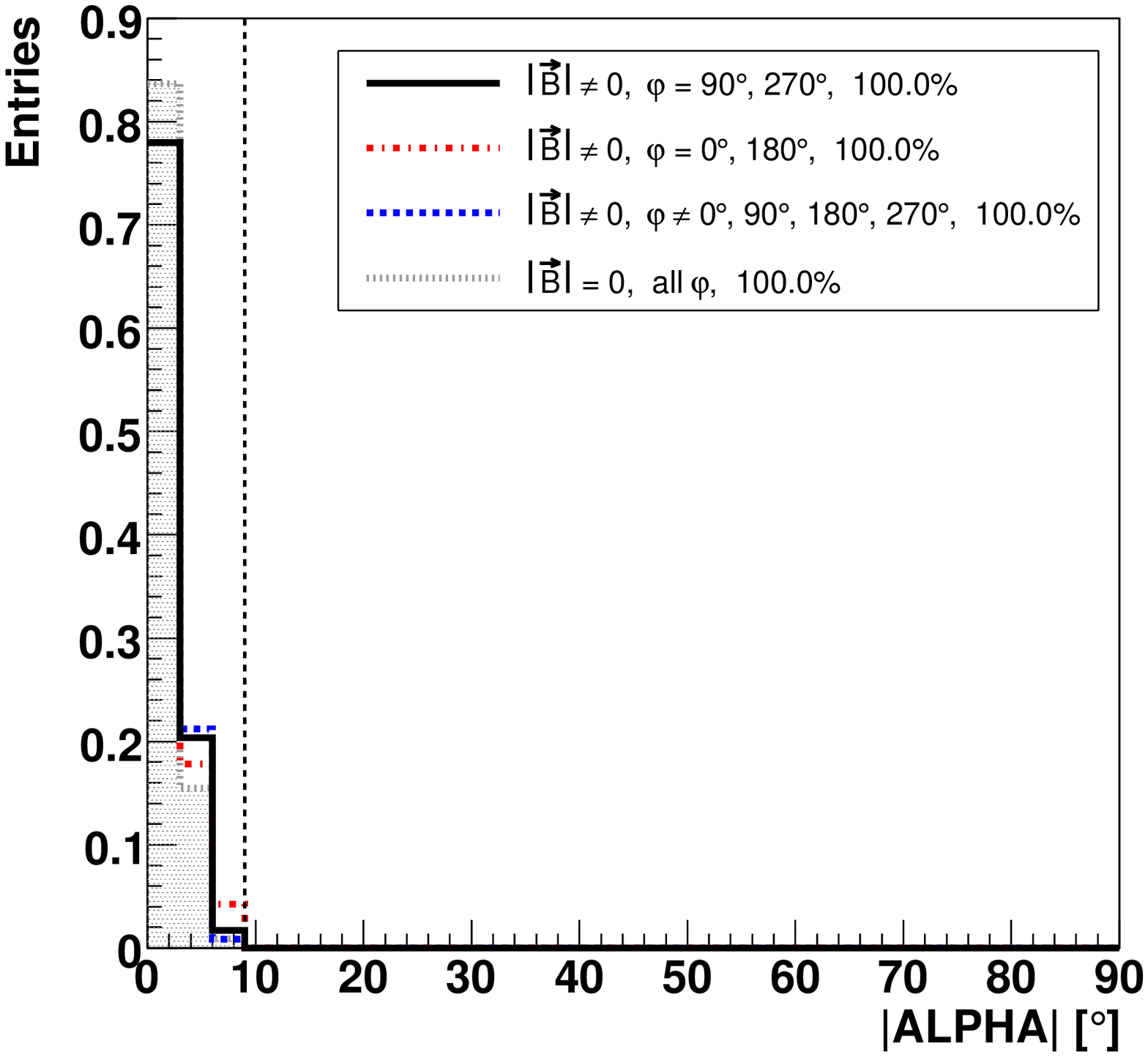}}\quad
  \subfigure[$\text{Az}=180^\circ$, $\text{IP}\approx 120\,\text{m}$, $\theta=87^\circ$.]{
    \includegraphics[scale=.33]{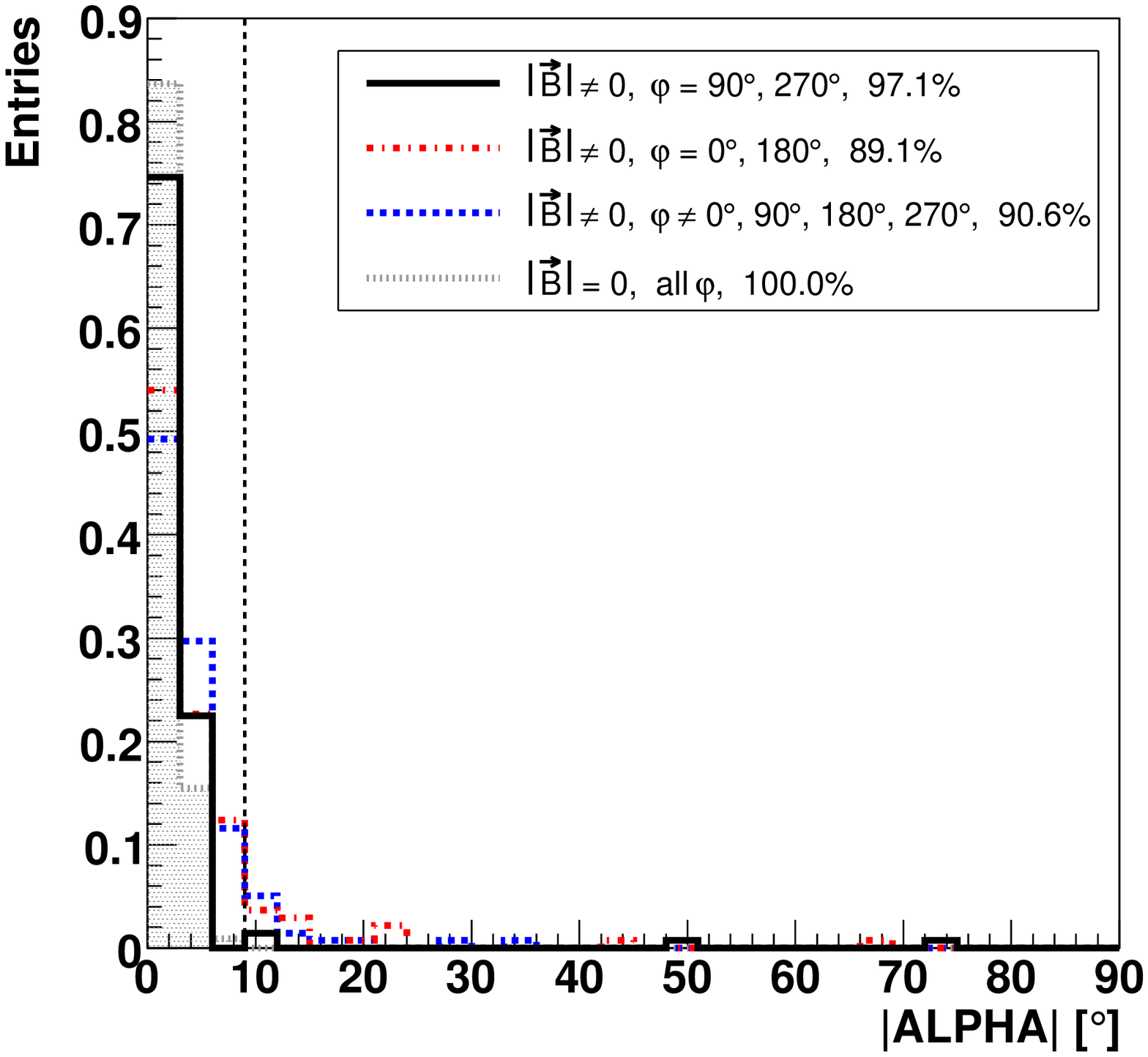}}
  \caption{As figure \ref{figure8}, but for 1\,TeV
    $\gamma$-rays, $\text{ZA}=40^\circ$, $\text{Az}=0^\circ$ and $180^\circ$.}
  \label{figure10}
\end{figure}

\begin{figure}[!ht] \centering
  \subfigure[$\text{IP}\approx 40\,\text{m}$, $\varphi = 0^\circ\dots 330^\circ$.]{
    \includegraphics[scale=.31]{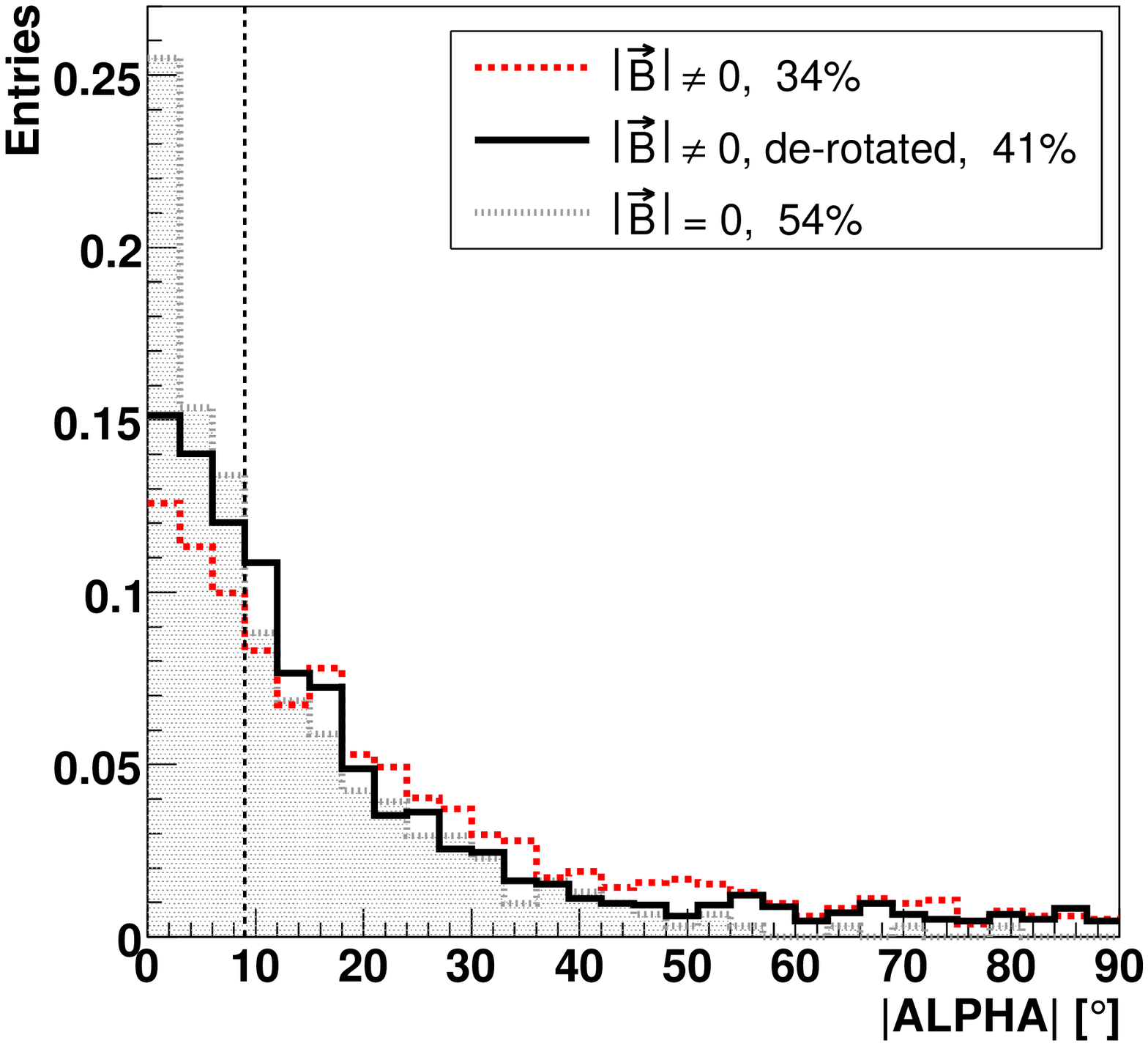}}\quad
  \subfigure[$\text{IP}\approx 120\,\text{m}$, $\varphi = 0^\circ\dots 330^\circ$.]{
    \includegraphics[scale=.31]{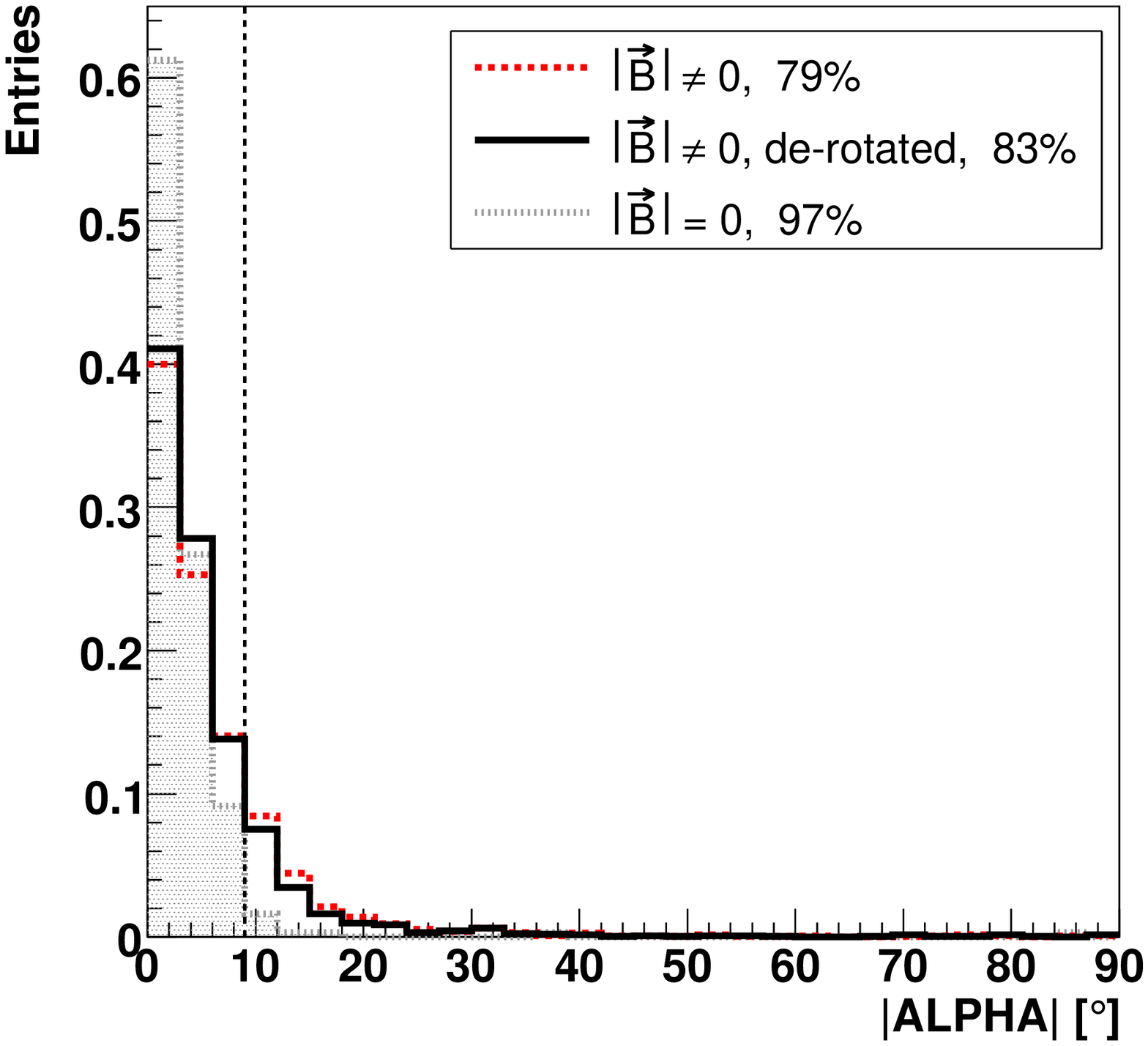}}\\
  \subfigure[$\text{IP}\approx 40\,\text{m}$, $\varphi \neq 0^\circ$ and $180^\circ$.]{
    \includegraphics[scale=.31]{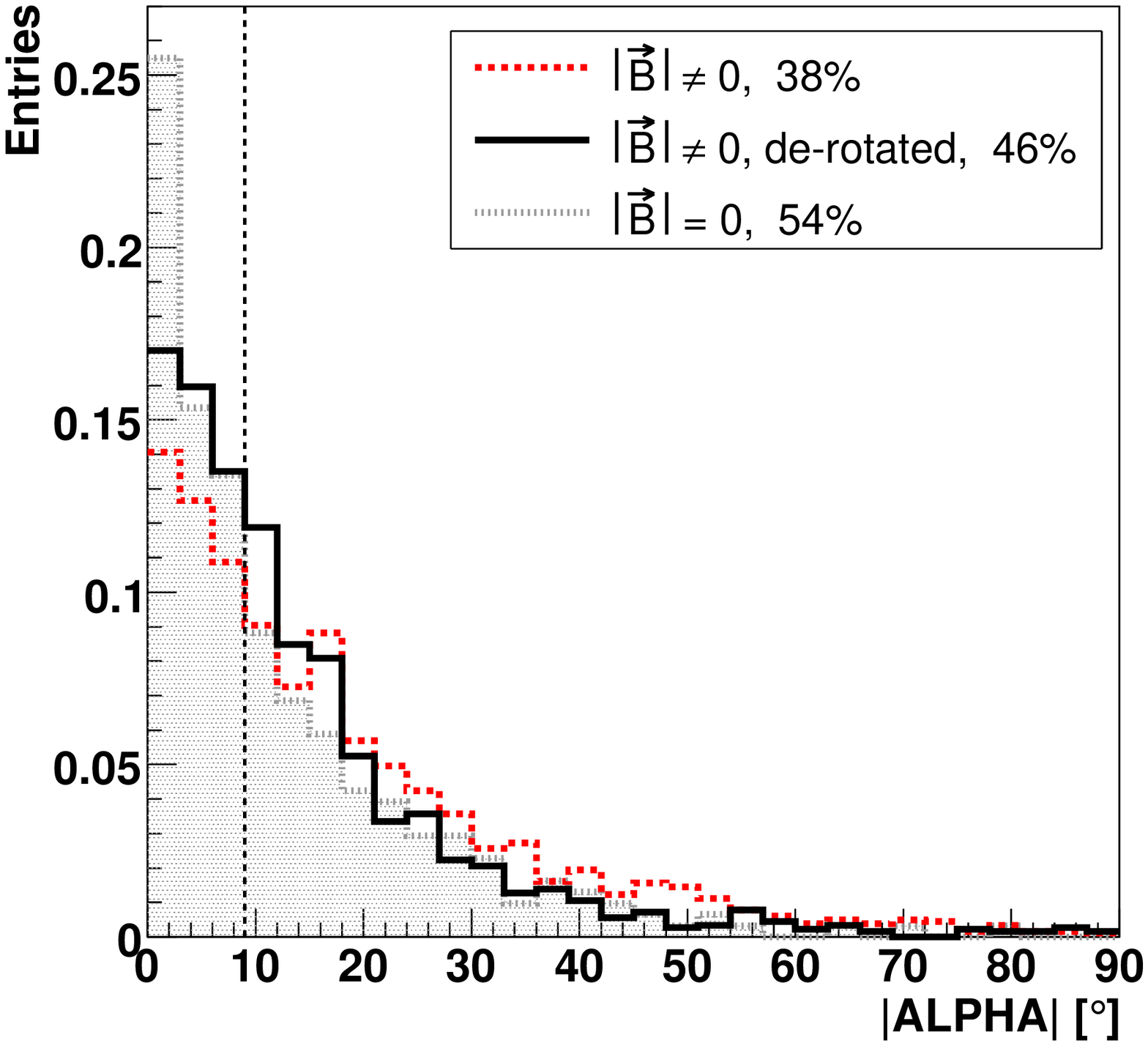}}\quad
  \subfigure[$\text{IP}\approx 120\,\text{m}$, $\varphi \neq 0^\circ$ and $180^\circ$.]{
    \includegraphics[scale=.31]{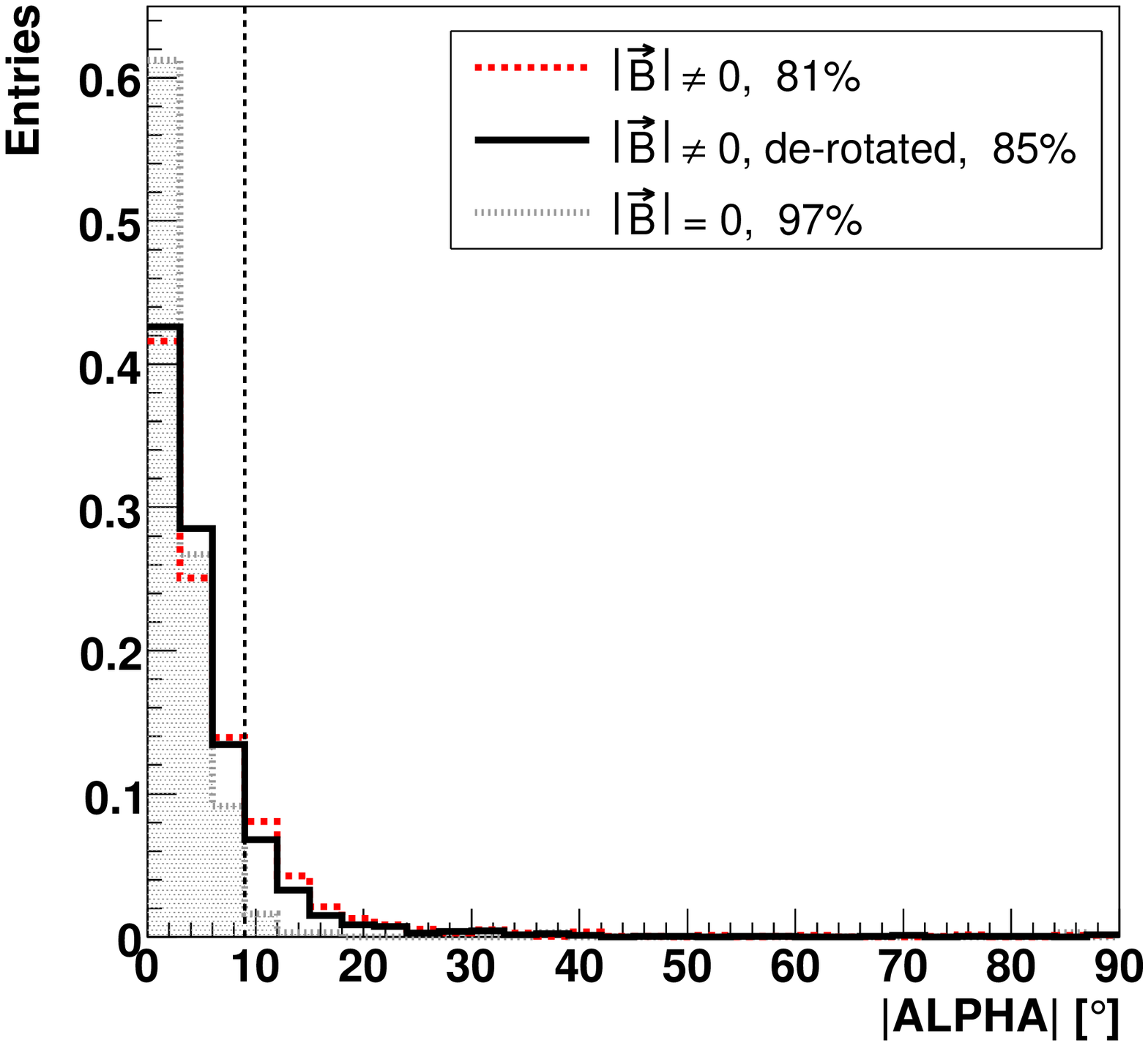}}\\
  \subfigure[$\text{IP}\approx 40\,\text{m}$, $\varphi = 0^\circ$ and $180^\circ$.]{
    \includegraphics[scale=.31]{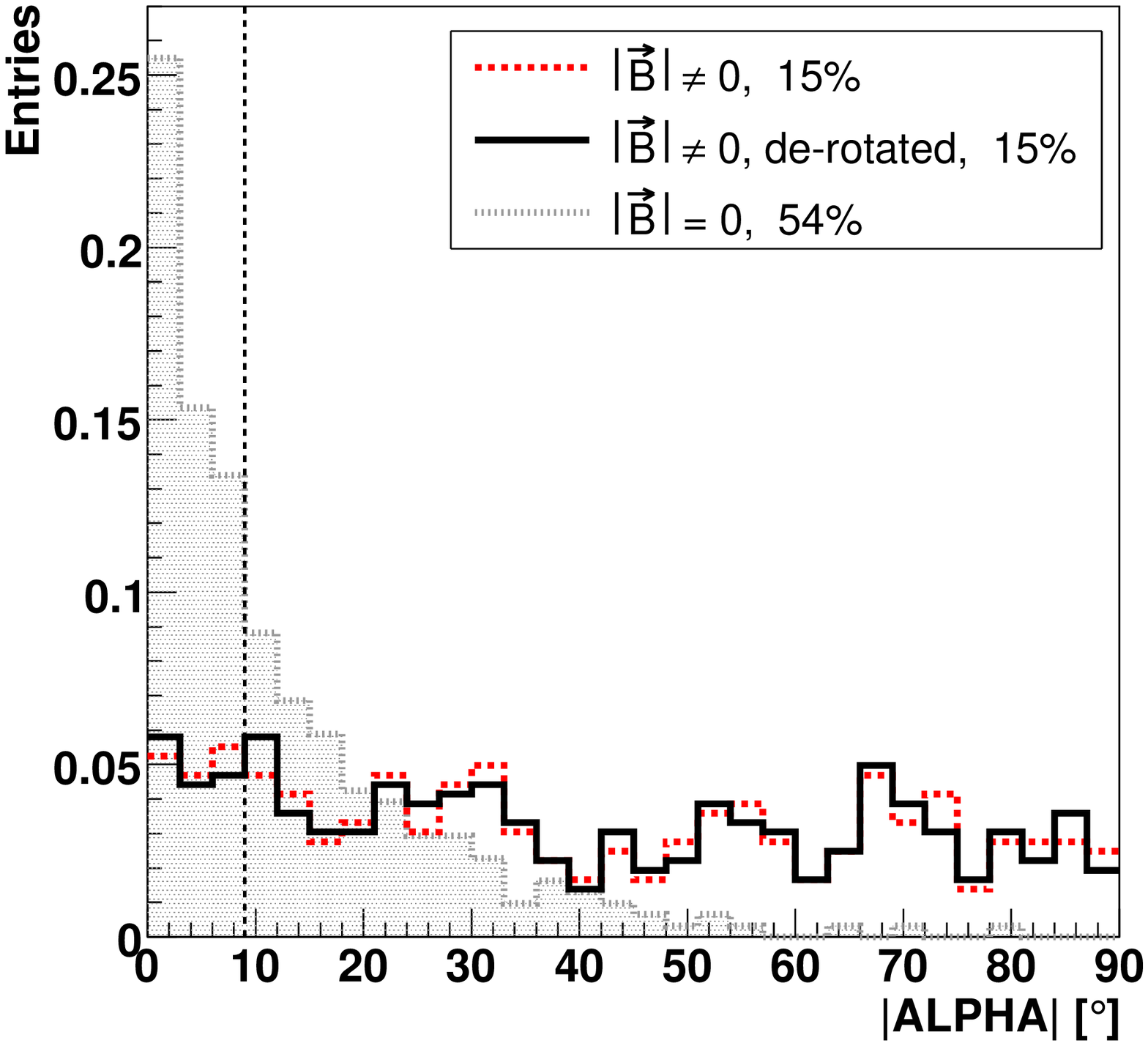}}\quad
  \subfigure[$\text{IP}\approx 120\,\text{m}$, $\varphi = 0^\circ$ and $180^\circ$.]{
    \includegraphics[scale=.31]{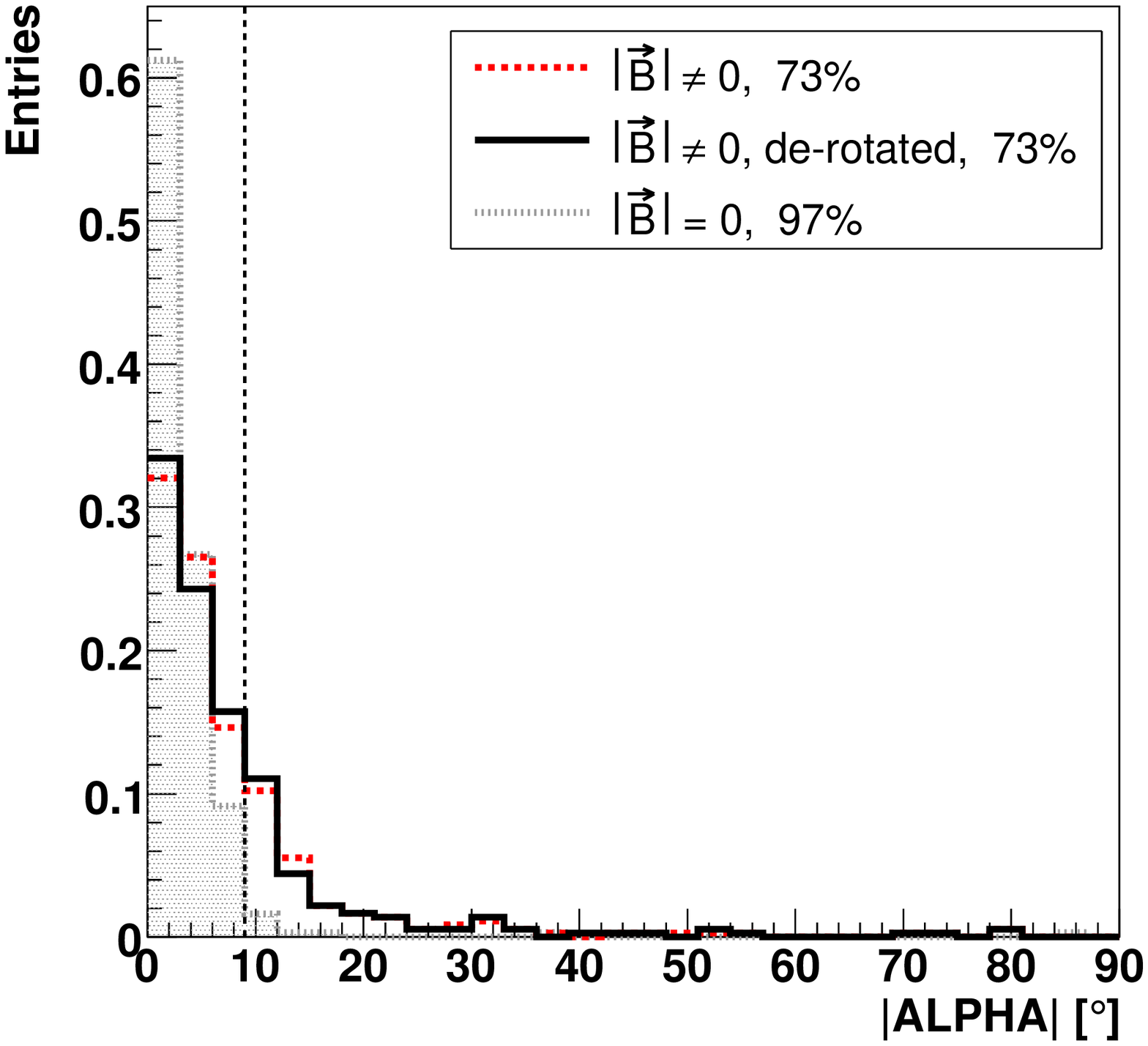}}
  \caption{Normalised distributions of the image parameter ALPHA for 450\,GeV
    $\gamma$-rays, $\text{ZA}=40^\circ$ and $\text{Az}=180^\circ$,
    40\,m ((a), (c) and (e)) and 120\,m impact parameter ((b), (d) and
    (f)). The distributions indicated by dotted lines were
    obtained without GF and the ones indicated by dashed lines for enabled GF in
    the MC simulation. For the distributions indicated by solid lines the shower images were
    de-rotated (see text for more details).}
  \label{figure11}
\end{figure}

Figures \ref{figure8}\,-\,\ref{figure10} show the ALPHA distributions (normalised to the number of entries) for
$\gamma$-ray energies of 50\,GeV, 450\,GeV and 1\,TeV.
Showers recorded at 40\,m and 120\,m impact parameter
were considered together with the different possible configurations:
the distributions indicated by dash-dotted lines correspond to
$\varphi=0^\circ$ and $180^\circ$ (as defined in figure \ref{figure2} (a)),
where the connecting line between the shower axis
and the telescope optical axis is parallel to the north-south direction.
This configuration corresponds to shower images which are oriented parallel to the direction of
the GF in the camera. The distributions indicated by
solid lines were obtained for $\varphi=90^\circ$ and
$270^\circ$, where the connecting line between the shower axis
and the telescope optical axis is parallel to the east-west direction.
In this case the shower images are on average not rotated and oriented normal to
the direction of the GF in the camera.
The ALPHA distributions indicated by dashed lines belong to intermediate telescope
positions $\varphi = 30^\circ$, $60^\circ$, $120^\circ$, $150^\circ$, $210^\circ$,
$240^\circ$, $300^\circ$ and $330^\circ$. The corresponding ALPHA
distributions obtained without GF in the MC simulation are also plotted
(dotted lines). The percentage of events with $|\text{ALPHA}|\leq 9^\circ$ is given in the legend.
The cut is indicated by the vertical dotted line.\\
It can be seen that for configurations where the connecting line between
the telescope and the shower axis is parallel to the north-south direction
(parallel to the direction of the GF) the corresponding ALPHA distributions
(red histograms) can be significantly broadened although the corresponding shower images
are on average not rotated. However, the ALPHA distribution for the opposite
configuration (connecting line between the telescope and the shower axis
parallel to the east-west direction) are stronger peaked at low values
due to the influence of the GF (compare figure \ref{figure9} (a) and (b)).
The remaining configurations always lead to broadened ALPHA distributions due
to the rotation of the shower images with a preferential direction.\\
In conclusion it can be stated that the influence of the GF can significantly
degrade the orientation discrimination of shower images. It is evident that for some
configurations discussed above the $\gamma$-ray signal cannot be recovered
by de-rotating the shower images.\\
Figure \ref{figure11} shows the ALPHA distributions for 450\,GeV $\gamma$-rays, $\text{ZA}=40^\circ$,
$\text{Az}=180^\circ$ and 40\,m as well as 120\,m impact parameter. The ALPHA distributions
obtained without GF (dotted line) are shown together with the distributions obtained for enabled GF
(dashed line) and the ones obtained after de-rotation of the
shower images (solid line). It is possible to correct
for GF effects by de-rotating the shower images.
However, if all telescope positions are taken into
account ($\varphi=0^\circ\dots 330^\circ$) the improvement in terms of the percentage of events with
$|\text{ALPHA}|\leq 9^\circ$ is less than 8\,\% (figure
\ref{figure11} (a) and (b)). Ignoring the most
unfavourable configurations with respect to the influence of the GF
($\varphi\neq 0^\circ$ and $180^\circ$) 
results in ALPHA distributions which are stronger peaked at small values (figure
\ref{figure11} (c) and (d)). 
As expected, for the most unfavourable configurations ($\varphi=
0^\circ$ and $180^\circ$) the $\gamma$-ray signal cannot be recovered
by de-rotation. The corresponding images are not
rotated but their angular distribution is broadened (figure
\ref{figure11} (e) and (f)).\\
Given that the energy, the rotation angle and the impact parameter is well
known in MC, the amount of recovered real $\gamma$-ray showers from observational data
by de-rotation is expected to be lower. Both the energy and
the impact parameter have to be estimated and are thus known less precisely.
To be efficient, the de-rotation of the shower images requires a precise knowledge of the impact parameter
and the information on the energy of the $\gamma$-ray candidates from
observational data. Moreover, we focused on intermediate $\gamma$-ray energies where the rotation angle is
large (figure \ref{figure5} (d)) and the spread of the ALPHA distribution is rather low.
At lower energies than those
considered here the recovery of the $\gamma$-ray signal is even less
efficient \cite{com0701}. This is also the case for 1\,TeV $\gamma$-rays,
where the ALPHA distribution is stronger peaked at small values and the
rotation angle is smaller (figure
\ref{figure5}). Because of the relatively
poor knowledge of the impact parameter in case of real shower images the
improvement in sensitivity by de-rotation of the shower images is
expected to be below 10\,\%.

\clearpage

\clearpage

\clearpage

\clearpage

\subsection{GF Effects on the DISP-reconstructed $\gamma$-ray Arrival Direction}

A DISP analysis of the MC $\gamma$-ray shower images was performed to study
the GF effects on the reconstructed arrival directions.

\begin{figure}[!h]
  \subfigure[$\text{Az}=0^\circ$, $\theta=12^\circ$.]{
    \includegraphics[scale=.33]{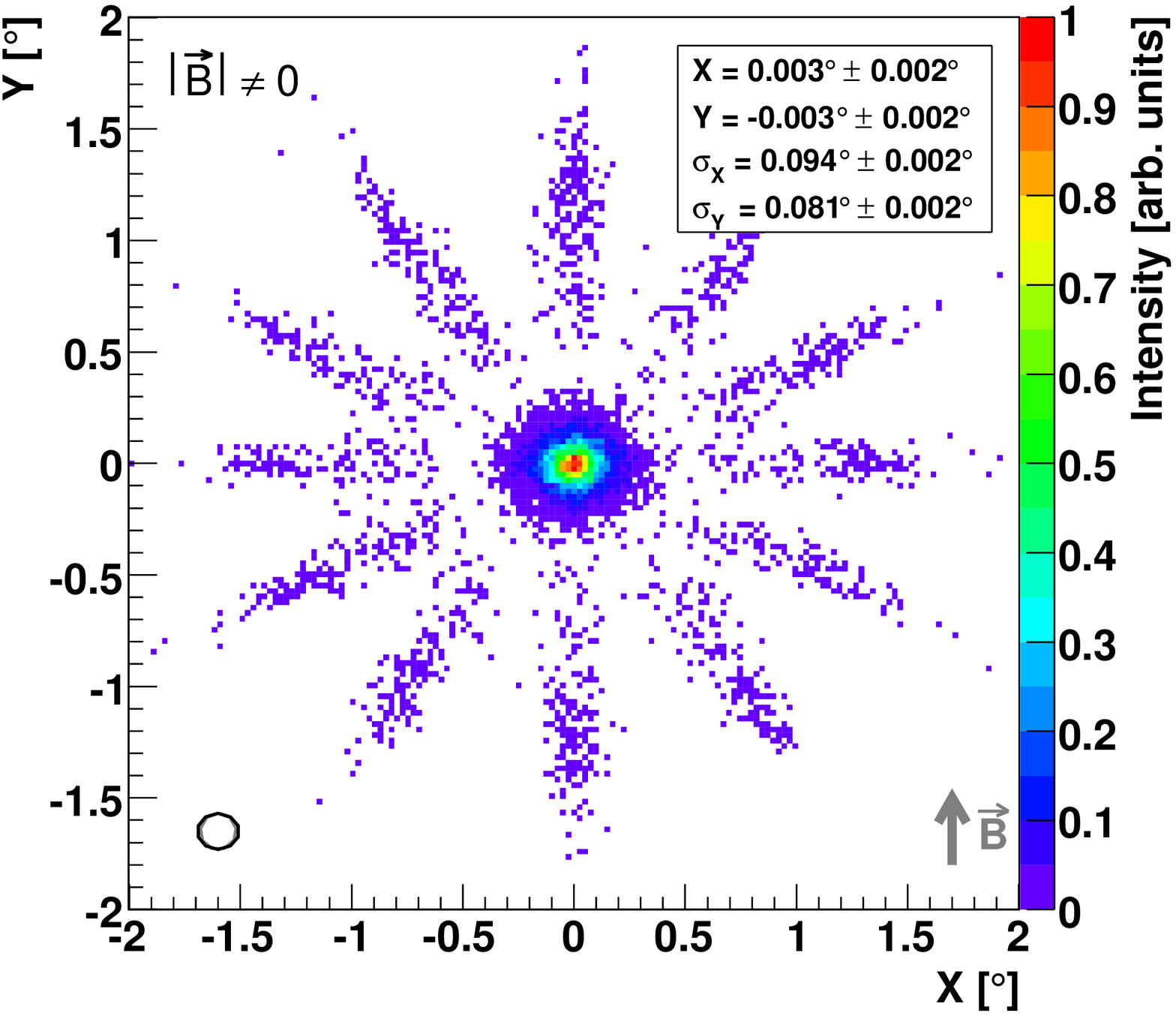}}\quad
  \subfigure[$\text{Az}=180^\circ$, $\theta=87^\circ$.]{
    \includegraphics[scale=.33]{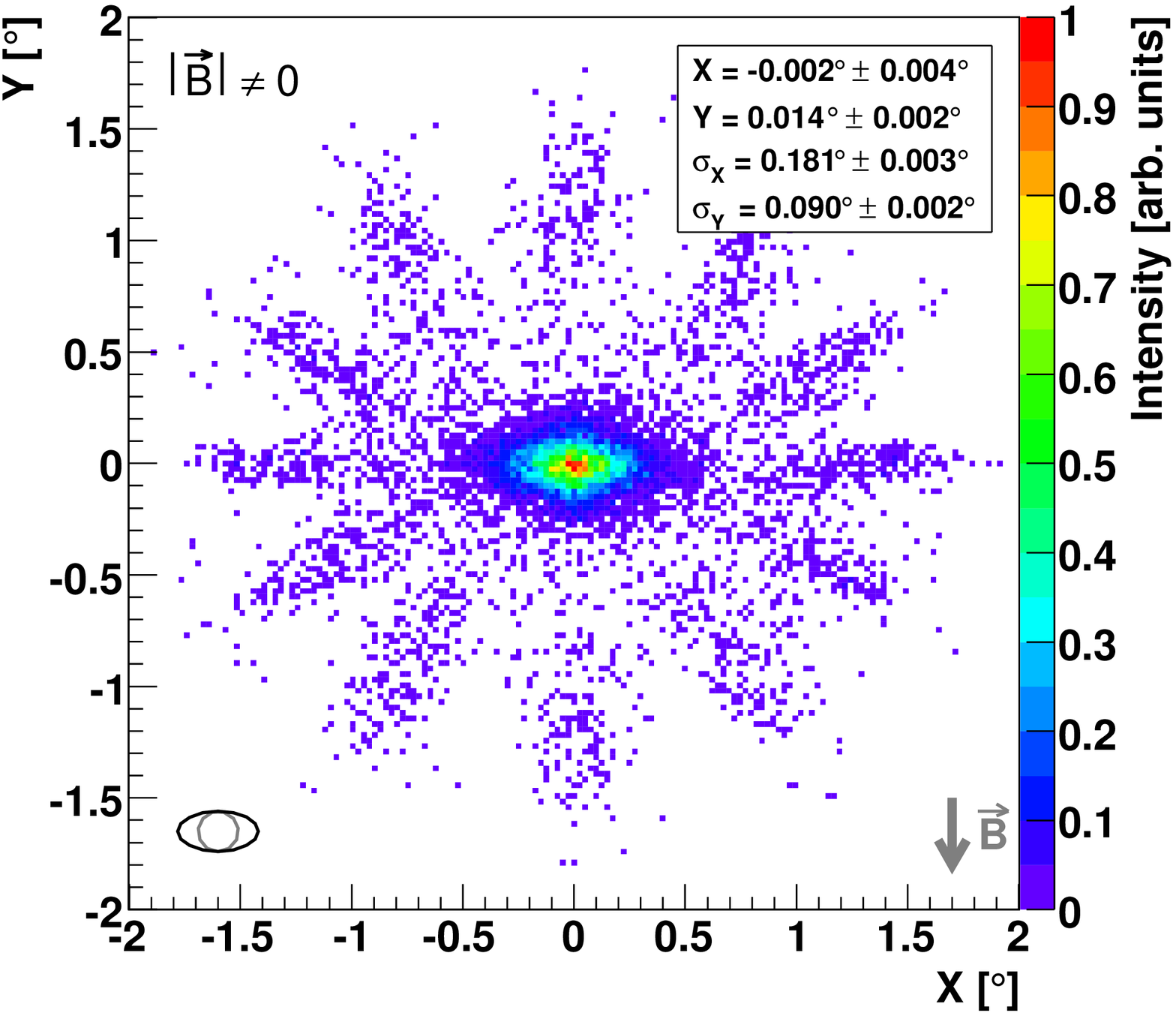}}
  \caption{DISP-reconstructed arrival directions for 450\,GeV $\gamma$-rays,
    impact parameters between 60\,m and 140\,m, $\text{ZA}=40^\circ$,
    $\text{Az}=0^\circ$ and $180^\circ$, respectively (see text for more details).}
  \label{figure12}
\end{figure}

For the optimisation of the second-order polynomials in equation (\ref{eq:disp}),
dedicated MC $\gamma$-ray samples with continuous impact parameter
distribution between 0\,m and 500\,m were produced.
The MC samples were produced for the same $\gamma$-ray energies, ZAs and image cleaning levels
as the MC data used for the preceding studies.
The EAS core location was
randomly placed somewhere in a circle on the plane perpendicular to the
direction of the EAS. Also, the MC samples were produced without GF,
thus only for $\text{Az}=0^\circ$. In this way the results from the DISP
method obtained for different telescope pointing directions are comparable
since the DISP polynomials themselves are not subject to GF effects.\\
Figure \ref{figure12} shows the DISP-reconstructed arrival directions for 450\,GeV $\gamma$-rays,
impact parameters between 60\,m and 140\,m, $\text{ZA}=40^\circ$, $\text{Az}=0^\circ$ and $180^\circ$.
The projected direction of the GF is
indicated in the lower right part of the figures, and the ellipticity of the
distributions of DISP-reconstructed arrival directions
is shown in the lower left part of the figures.
The semi-minor and the semi-major axis of the ellipse correspond to the sigma
of a Gaussian fit to the distributions using bands of $\Delta_{X,Y}=\pm 0.035^\circ$
parallel and perpendicular to the projected direction of the GF.
The size of the bands was arbitrarily chosen, but it is not critical for the
result. Within errors, the relative difference of the results obtained from the two
orientations is independent of the size of the bands.
The result from the Gaussian fit is also shown in the legend.\\
As can be seen from the figures the distributions appear to be significantly elongated
perpendicular to the projected direction of the GF,
while the peak of the DISP distribution is always centred at
the nominal source position (camera centre). The extent of the elongation
depends on the angle $\theta$ between the shower axis and
the direction of the GF. The GF effects on the DISP method thus
result in a degradation of the sky maps in a way that a point-like
$\gamma$-ray source appears to be extended, i.e. the
$\gamma$-ray PSF is degraded.
The star-shaped appearance of the DISP distributions arises from events with
wrong head-tail assignment. The false head-tail assignment cannot be
attributed to GF effects since it occurs also for the favourable telescope
pointing direction (figure \ref{figure12} (a)).
This illustrates the basic limitation of a single
telescope to properly reconstruct the true source position. In case of a single telescope
the DISP method has to rely on the shower asymmetry along the major axis of
the shower image (see section \ref{sec:analysis}).

\begin{figure}[!h] \centering
  \subfigure[$\text{Az}=0^\circ$, $\text{ZA}=40^\circ$, $\theta=12^\circ$.]{
    \includegraphics[scale=.33]{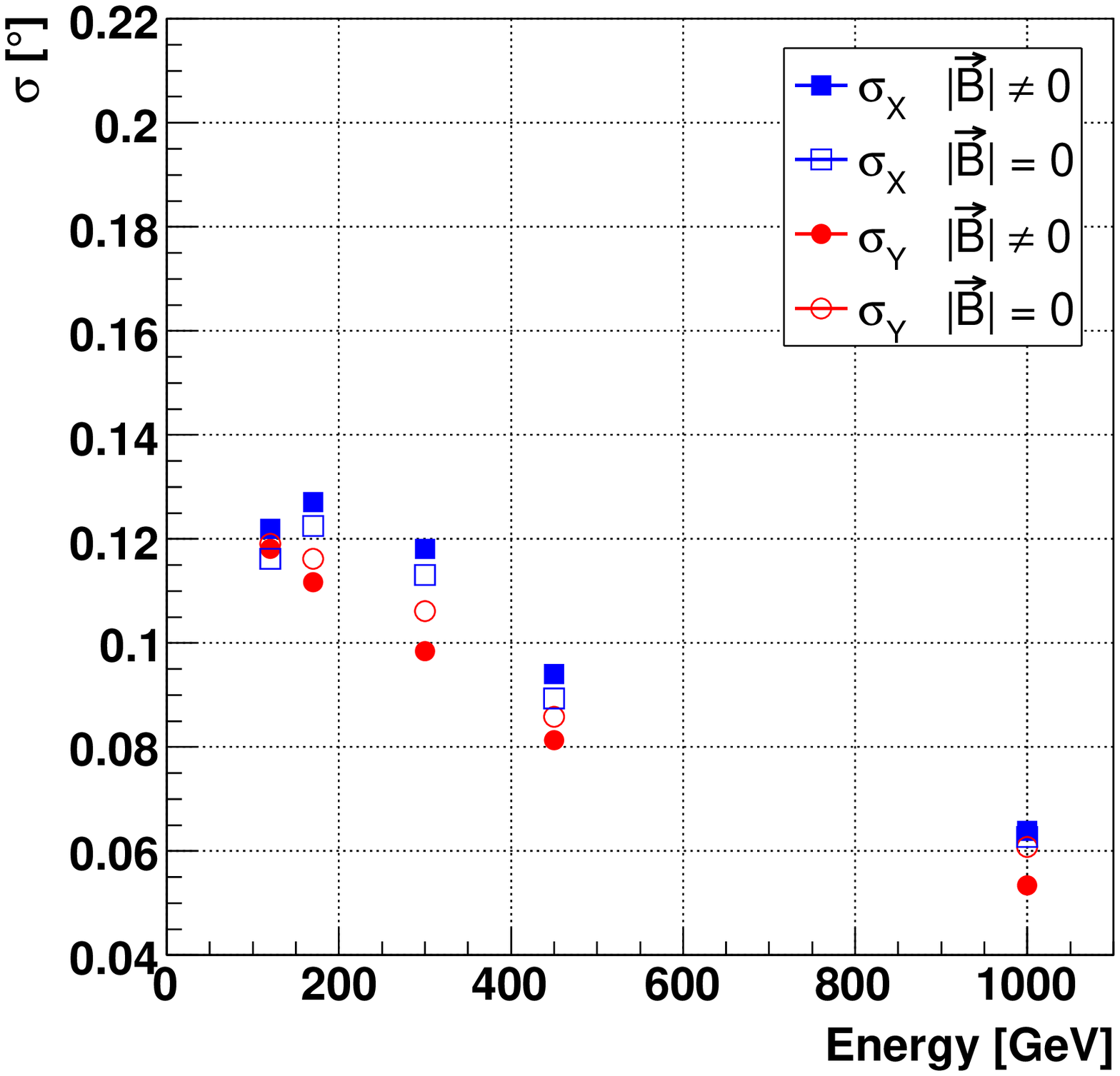}}\quad
  \subfigure[$\text{Az}=180^\circ$, $\text{ZA}=40^\circ$, $\theta=87^\circ$.]{
    \includegraphics[scale=.33]{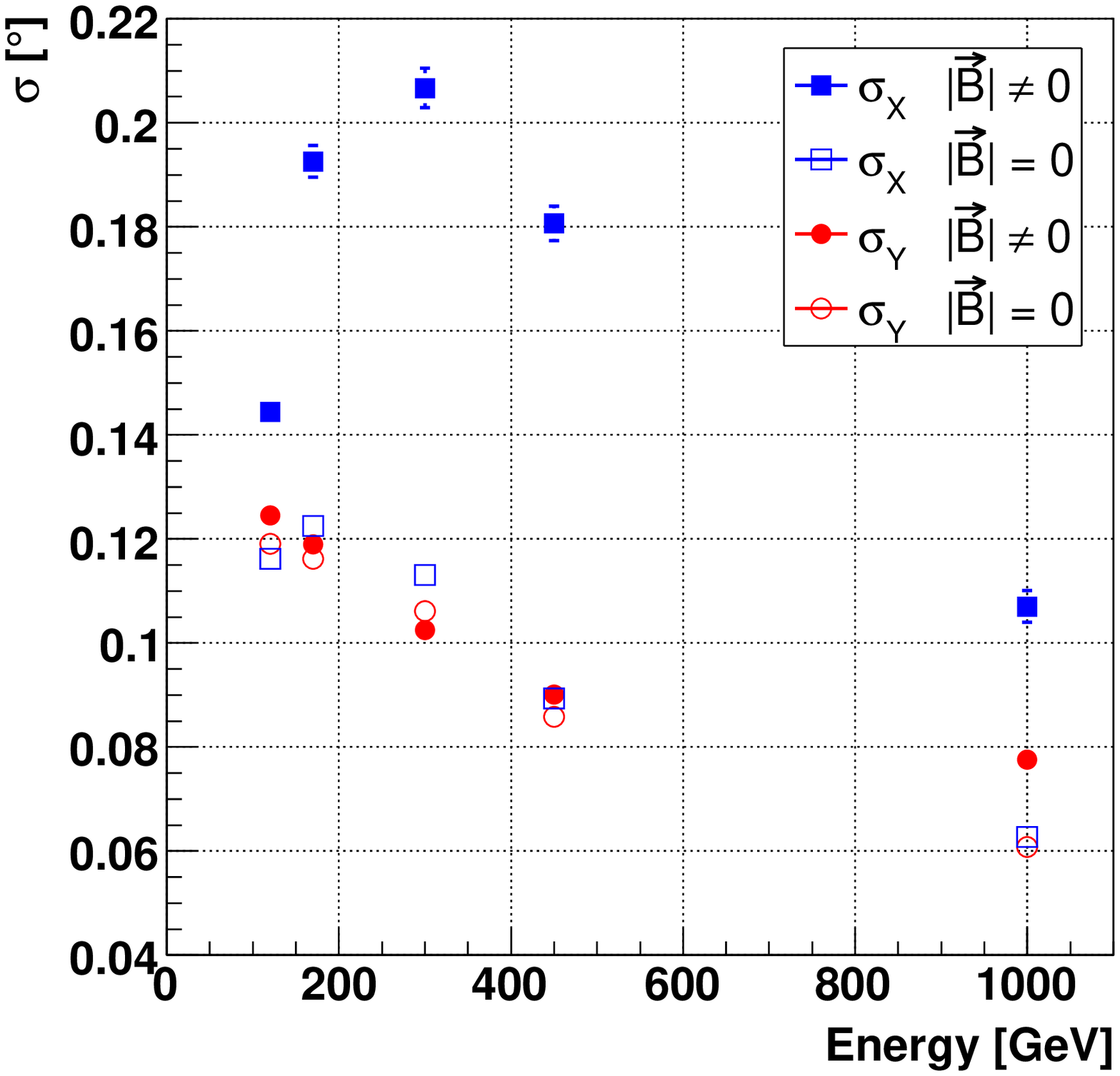}}
  \caption{Lateral and longitudinal spread of the DISP distribution
    versus $\gamma$-ray energy for impact parameters between 60\,m
    and 140\,m, $\text{ZA}=40^\circ$, $\text{Az}=0^\circ$ and $180^\circ$,
    respectively (see text for more details).}
  \label{figure13}
\end{figure}

Figure \ref{figure13} shows the
lateral and longitudinal spread versus $\gamma$-ray energy for
different orientations of the telescope. The spread is defined as the sigma
of a Gaussian fit to the DISP distribution using bands of $\Delta_{X,Y}=\pm 0.035^\circ$
parallel and normal to the projected direction of the GF in the
camera. Impact parameters between 60\,m and 140\,m were considered.
The figures clearly show that, compared to the case of disabled GF,
the spread of the DISP distribution
increases significantly for an unfavourable telescope orientation
(large angle $\theta$). The maximum spread occurs
always perpendicular to the direction of the GF in the camera.
Therefore, depending on the orientation of an EAS with respect to the
telescope and the impact parameter, the DISP-reconstructed incoming direction of the
corresponding primary $\gamma$-ray has a large uncertainty.

\subsection{GF Effects on the Energy Reconstruction of $\gamma$-ray Images}

It was previously discussed that the influence of the GF on the shower development
affects also the energy reconstruction and the trigger efficiency for
$\gamma$-rays. 

\begin{figure}[!h] \centering
  \subfigure[$\text{ZA}=0^\circ$.]{
    \includegraphics[scale=.32]{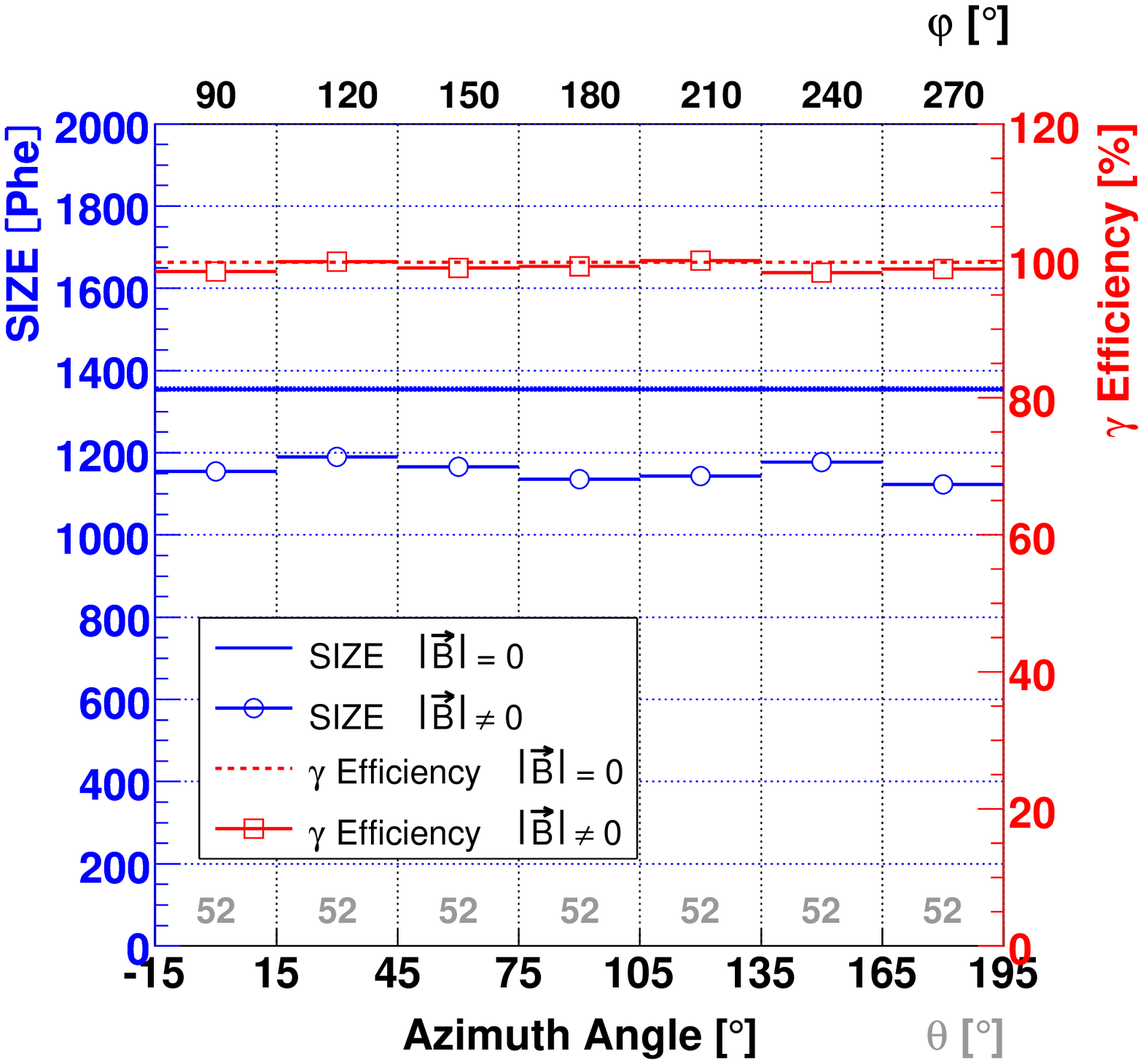}}\qquad
  \subfigure[$\text{ZA}=20^\circ$.]{
    \includegraphics[scale=.32]{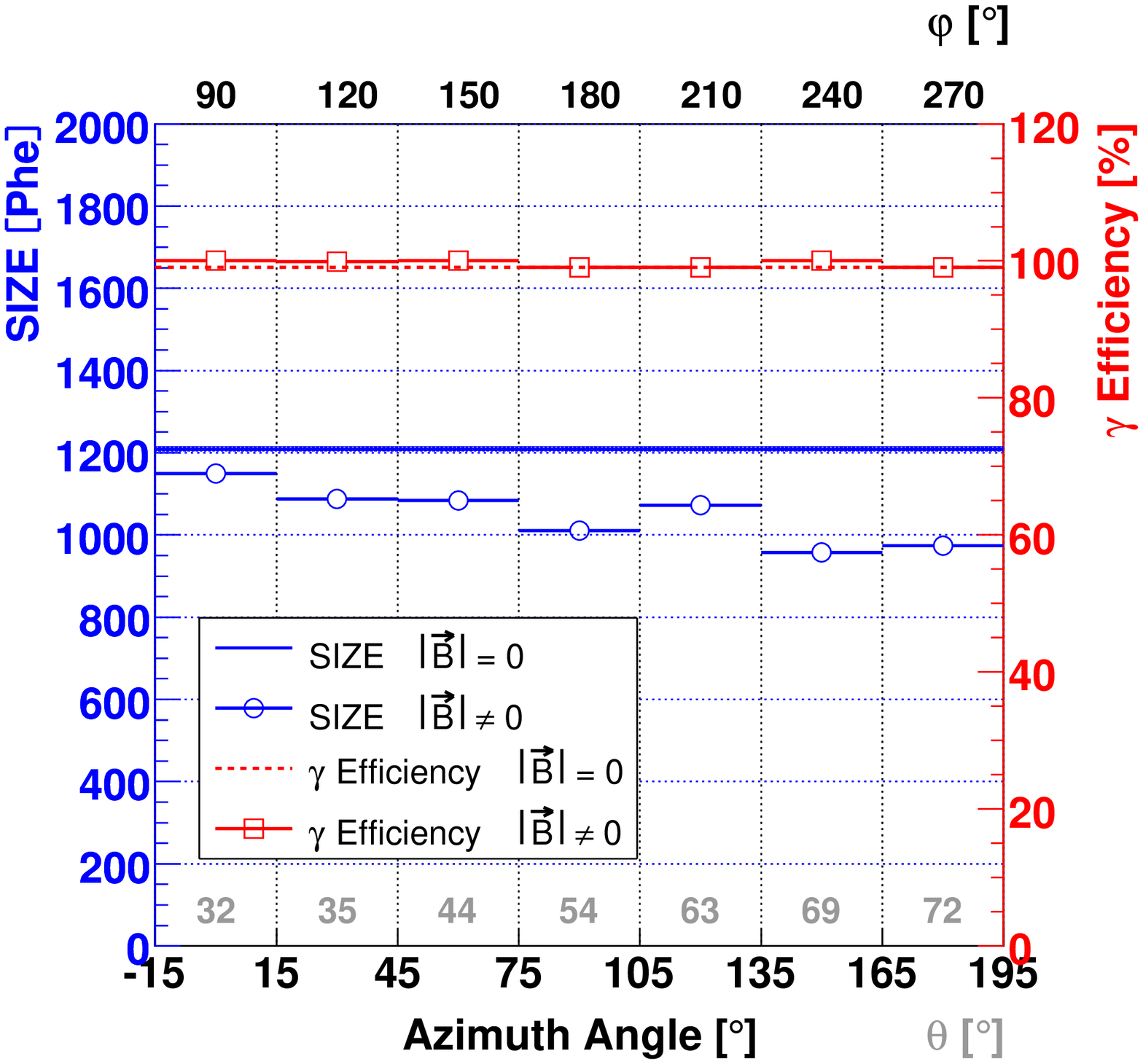}}
 \subfigure[$\text{ZA}=40^\circ$.]{
    \includegraphics[scale=.32]{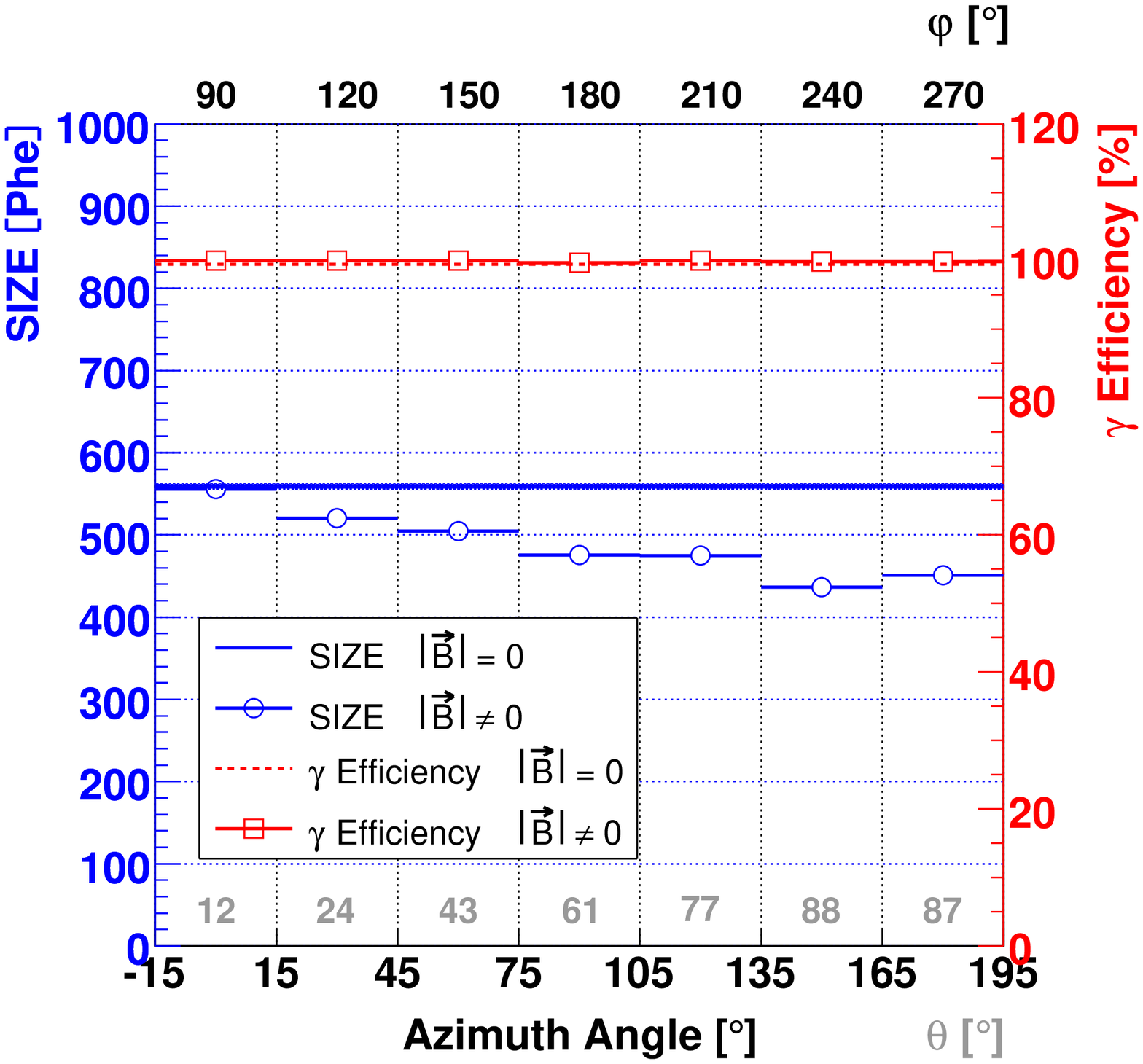}}\qquad
  \subfigure[$\text{ZA}=60^\circ$.]{
    \includegraphics[scale=.32]{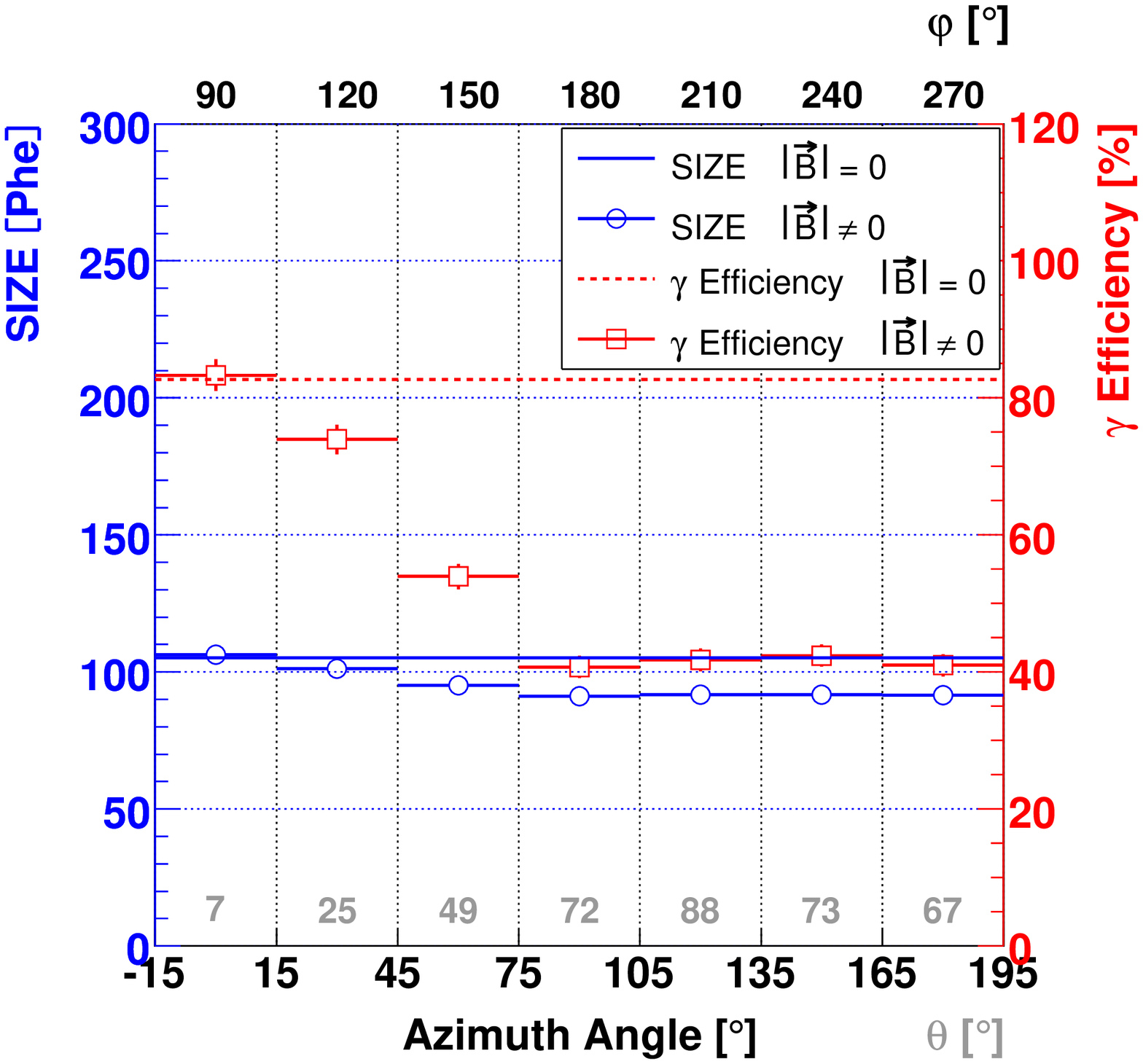}}
  \caption{The mean of the image parameter SIZE and
    the $\gamma$ efficiency versus Az angle for 450\,GeV $\gamma$-rays and ZA between
    $0^\circ$ and $60^\circ$. The telescope is
    always situated at angles $\varphi=\text{Az} + 90^\circ$ and
    $120\,\text{m}$ impact parameter.}
    \label{figure14}
\end{figure}

The east-west separation of electrons and positrons in EAS due to the
GF modifies the Cherenkov distribution on the ground such that the
reconstructed Cherenkov light, i.e. the integrated light content
of the shower images is reduced for unfavourable shower orientations with
respect to the direction of the GF.\\
Figure \ref{figure14}
shows the average reconstructed light content of shower images
together with the $\gamma$ efficiency versus Az angle
for 450\,GeV energy $\gamma$-rays and ZAs between $0^\circ$ and
$180^\circ$. To compare equivalent configurations (see figure
\ref{figure2} (a)) the telescope is always situated at angles $\varphi=\text{Az} +
90^\circ$ and the impact
parameter was set to 120\,m. The angle $\theta$
between the direction of the EAS and the GF is given on top of the abscissa.
The $\gamma$ efficiency is defined as the ratio of the number of $\gamma$-ray
showers surviving the trigger and the image cleaning to the number of generated
$\gamma$-rays. From figure \ref{figure14} (d)
it can be seen that the $\gamma$ efficiency varies by up to
50\,\%. For showers close to the trigger threshold the Cherenkov light
distribution on the ground
can be thinned out such that most of the events do not survive the trigger level.
This occurs only for showers close to the energy threshold of
the telescope, which is ZA dependent.\\
For some telescope pointing directions the total reconstructed integrated light of
shower images can be reduced by up to $\sim 20\,\%$.
This is not only the case for low energies but also for
TeV $\gamma$-rays. Consequently, if GF effects are
not taken into account the energy of $\gamma$-ray
candidates from observational data will be systematically underestimated
whereas the $\gamma$ efficiency will be overestimated. Both effects degrade
the determination of the flux from a $\gamma$-ray source if they are not properly
taken into account in the MC simulation.

\section{Conclusions}

The results from the MC studies show that the GF can significantly affect both
the shape and the orientation of shower images recorded with an IACT like MAGIC. Therefore,
the orientation discrimination of $\gamma$-rays against unwanted (hadronic)
background can be significantly degraded. It was demonstrated that the
de-rotation of the shower images does not help to recover the pointing
entirely. At most 10\,\% of the events can be recovered by de-rotation
requiring the knowledge of the impact parameter and energy of the $\gamma$-rays.\\
The influence of the GF also degrades the DISP-estimated arrival direction
of MC-generated $\gamma$-rays. Due to the
influence of the GF on the development of EAS the DISP distribution can be
significantly elongated perpendicular to the projection of the GF in the
camera. The quality of a sky map is degraded in a way that a point-like source appears
extended unless it is compared to a proper MC simulation taking into account
the trajectory of the source in the sky.
However, the peak of the DISP distribution is always centred at the
source position.\\
It was also shown that the influence of the GF on EAS can significantly
affect the energy reconstruction and the trigger
efficiency for $\gamma$-rays. If this effect is not taken into account,
the energy of $\gamma$-ray candidates from observational data will be
systematically underestimated (up to $\sim 20\,\%$ effect).
For low energies close to the analysis threshold ($<100\,\text{GeV}$) the
$\gamma$ efficiency also depends on the position of the telescope in the
Cherenkov light pool \cite{com0701}. At higher energies ($\sim 300\,\text{GeV}$\,-\,$1\,\text{TeV}$),
the $\gamma$ efficiency is affected only at large ZA,
where the telescope threshold energy is significantly increased ($\lesssim 50\,\%$
effect).\\
It was demonstrated that the extent of the GF effects not only depends on the orientation of EAS
with respect to the direction of the GF but also on the position of the
telescope with respect to the EAS core location on ground.
Shower images are not only rotated away from the projected direction of the GF in the
telescope camera plane but can also be rotated towards it, contrary to what
was reported in \cite{cha9901}.\\
Altogether, GF effects on EAS affect the $\gamma$-ray sensitivity of an
IACT and the determination of the flux from a VHE $\gamma$-ray source.
Distinct MC data covering the same ZA and Az angle range as the observational data being
analysed are required to account for GF effects.\\
It is remarkable that the GF effects not only occur at very low energies but also
at high energies around 1\,TeV. The GF effects are rather pronounced at
$\gamma$-ray energies around 450\,GeV.
The reason for GF effects to occur at high energies is
presumably linked to a characteristic feature in the development of a
$\gamma$-ray induced EAS. The process of multiplication in EAS
continues until the average energy of the shower particles is insufficient
to further produce secondary particles in subsequent collisions. At this stage
of the shower development, the shower maximum is reached (largest number of secondary
particles) and the average energy of the secondaries is close to the so-called
critical energy of $\sim 100\,\text{MeV}$ \cite{bet34} below which secondary electrons and positrons
lose their energy predominantly through ionisation of air molecules \cite{gre56}.
At the shower maximum, the average energy of the secondary particles is independent of
the primary $\gamma$-ray energy and the GF has on average the same influence
on the secondary particles.\\
Apart from that, the average atmospheric depth at which the shower maximum
occurs increases logarithmically with increasing energy of the primary $\gamma$-ray \cite{gre56},
and therefore the track along which secondary electrons and
positrons suffer from Lorentz deflection increases, too.\\
Another point worthy of mentioning is the fact that the threshold energy for a
charged particle to emit Cherenkov light decreases with increasing atmospheric depth.
Hence, in high-energy EAS, even charged secondaries of lower energy suffering strong Lorentz deflection
may additionally contribute to the Cherenkov light pool on ground.\\
GF effects on the hadron induced background were not studied. It is
impossible to show the rotation effect using shower images from hadron
candidates of observational data, because they do not point to any source.
Also, possible GF effects on the hadron induced background
presumably do not degrade the background discrimination. Close to the
energy threshold the trigger efficiency for the hadronic background should be
reduced, too.

\section{Outlook}

Although this study focuses on GF effects relevant for a single IACT like
MAGIC the influence of the GF on the development of EAS is expected to degrade also
the performance of stereoscopic IACT arrays.
In a stereoscopic telescope system multiple telescopes view the same EAS.
The individual images are then combined to form a common event.
Therefore, stereoscopic systems allow for a three-dimensional reconstruction of the
shower axis resulting in an improved sensitivity.
The shower direction on the sky is estimated from the intersection point
of the major image axes in a composite field of view, on an event by event
basis \cite{feg97}. Due to the rotation of the individual shower images the
intersection point will be modified in a way that the source
direction in the sky is wrongly reconstructed.
Stereoscopic systems do not trigger homogeneously but preferably on EAS with
impact positions between the telescopes.
As the impact distance on ground between the
EAS and the individual telescopes of an array is different in the
majority of cases, the GF will deteriorate the orientation of the
individual shower images differently. As a result the performance of
stereoscopic IACT arrays is degraded by GF effects.
A detailed MC study on the influence of GF effects on the performance of
stereoscopic IACT systems is in preparation.\\
The intensity of the GF on the Earth' surface ranges from about
$20\,\mu\text{T}$ to about $70\,\mu\text{T}$ \cite{ngdc}.
It is therefore important to take into account the GF effects
for the site selection of future projects utilising the imaging air Cherenkov technique.
To minimise the influence of the GF on the detector performance it is
mandatory to select a site with a low absolute value of the GF. Hence, the
best-suited location would be close to the so-called South Atlantic
Anomaly, where the GF strength is minimal, amounting to about one half of
the value for the MAGIC telescope site.\\
It is difficult to study GF effects in observational data. The elevation effect on
shower images complicates such studies. Directions with strong magnetic field correspond to large
ZA and the sensitivity of an IACT changes as a function of
the ZA as a result of changing shower image characteristics due
to increasing air mass with increasing ZA \cite{feg97}.
There are several requirements a $\gamma$-ray source should fulfil
to be an appropriate candidate for GF studies
in observational data: it should be strong, preferably point-like, stable and it
should follow a trajectory corresponding to a large GF component normal
to the telescope pointing direction (figure \ref{figure1}
(a)).\\
Preliminary results from studies on GF effects in observational data taken with MAGIC were
already shown in \cite{com0701,com0702}. It was demonstrated that
the pointing resolution of MAGIC allows to
study GF effects in observational data even for a very
low component of the GF normal to the shower direction.
However, an extensive study on GF effects in observational data is in
progress.

\end{document}